\documentclass[12pt]{article}
\usepackage{graphicx, epsf}
\usepackage{amsmath,amssymb,amsfonts}

\newcommand{\be}{\begin{equation}}
\newcommand{\ee}{\end{equation}}
\newcommand{\bea}{\begin{eqnarray}}
\newcommand{\eea}{\end{eqnarray}}

\def\del{\partial}
\def\f{{\rm f}}
\def\D{{\rm D}}

\def\OWL{{\rm OWL}}
\def\CWL{{\rm CWL}}
\def\XSB{{\chi{\rm SB}}}
\def\XSR{{\chi{\rm SR}}}
\def\a{\alpha}

\def\d{\delta}

\def\L{\Lambda}
\def\r{\rho}
\def\s{\sigma}
\def\t{\tau}

\begin{document}

\begin{center}\ \\ 
\vspace{60pt}
{\Large {\bf Open Wilson Lines and Chiral Condensates \\[4mm]
in Thermal Holographic QCD}}\\ 
\vspace{60pt}
{\large Philip C. Argyres$^1$, 
Mohammad Edalati$^2$, 
Robert G. Leigh$^2$\\[2mm]
and Justin F. V\'azquez-Poritz$^3$}\\
\vspace{25pt}
{\it $^1$Department of Physics,
University of Cincinnati, Cincinnati OH 45221, USA}\\
{\tt argyres@physics.uc.edu}\\ [4mm]
{\it $^2$Department of Physics,
University of Illinois at Urbana-Champaign, Urbana IL 61801, USA}\\
{\tt edalati@illinois.edu}, {\tt rgleigh@illinois.edu}\\ [4mm]
{\it $^3$Physics Department,
New York City College of Technology\\
The City University of New York, Brooklyn NY 11201, USA}\\ 
{\tt jvazquez-poritz@citytech.cuny.edu}
\end{center}

\vspace{30pt}

\centerline{\bf Abstract}
\noindent 
We investigate various aspects of a proposal by Aharony and Kutasov \cite{ak0803} for the gravity dual of an open Wilson line in the Sakai-Sugimoto model or its non-compact version.  In particular, we use their proposal to determine the effect of finite temperature, as well as background electric and magnetic fields, on the chiral symmetry breaking order parameter.  We also generalize their prescription to more complicated worldsheets and identify the operators dual to such worldsheets.

\newpage
\tableofcontents
\addtocontents{toc}{\protect\setcounter{tocdepth}{3}}
\newpage

\section{Introduction}

In \cite{ak0803} Aharony and Kutasov gave a prescription for computing correlators of open Wilson line (OWL) operators in the Sakai-Sugimoto model and its non-compact version.  These OWLs are interesting since their vevs are order parameters for chiral symmetry breaking ($\XSB$).  Recall that the Sakai-Sugimoto model \cite{ss0412} is holographically dual to a gauge theory, known as holographic QCD, which shares many dynamical features with QCD-- namely, confinement and $\XSB$.  The non-compact Sakai-Sugimoto model is holographically dual to a theory which we will call holographic Nambu-Jona-Lasinio (NJL) which, like the usual NJL model, has $\XSB$ \cite{ahjk0604}.  We review some details of these models below.

In this paper, we use the prescription of \cite{ak0803} to probe how the $\XSB$ order parameters in holographic QCD and NJL depend on temperature, background electric and magnetic fields, and various parameters defining the OWL operator.  Since confinement is not our main interest, we mostly focus on holographic NJL, where confinement is turned off.  Most of our results are qualitatively unchanged if we consider, instead, holographic QCD. This is analogous to the case in field theory where, when it comes to the analysis of chiral symmetry breaking, the NJL model shows qualitatively similar behavior to QCD. 

We will first give a brief review of the string models for holographic QCD and NJL, and for the proposals for $\XSB$ order parameters in these models. The Sakai-Sugimoto model \cite{ss0412} is an intersecting brane model made of $N_c$ ``color'' $\D4$-branes extended in the $(x^0 x^1 x^2 x^3 x^4)$-directions, with $x^4$ being a Scherk-Schwarz circle of radius $R_0$ intersecting $N_\f$ $\D8$-branes and $N_\f$ $\overline{\D8}$-branes at two $(3+1)$-dimensional intersections.  The ``flavor'' $\D8$- and $\overline{\D8}$-branes are located at the antipodal points on the $x^4$-circle.  Imposing anti-periodic boundary conditions for fermions around the $x^4$-circle leaves the $U(N_c)$ gauge boson modes of the $4-4$ strings massless, but gives mass to their fermionic and scalar modes.  There are also massless Weyl fermion modes localized at the D4-D8 and D4-$\overline{\D8}$ intersections, denoted $\psi_L$ and $\psi_R$, which come from the Ramond-Ramond sector of the $4$-$8$ and $4$-$\overline 8$ strings, and which transform as $(\bf{N_c}, \bf{N_\f}, \bf1)$ and $(\bf{N_c},\bf1,\bf{N_\f})$ of $U(N_c)\times U(N_\f)\times U(N_\f)$, respectively.  The $U(N_\f) \times U(N_\f)$ gauge symmetry of the flavor branes is the chiral symmetry for these Weyl fermions.

The low energy theory on the color branes is a $(4+1)$-dimensional $U(N_c)$ gauge theory with a dimensionful 't Hooft coupling $\lambda_5$. The dynamics of the theory at the intersections are governed by the dimensionless effective 't Hooft coupling $\lambda_4= \lambda_5/R_0$.  At weak coupling ($\lambda_4 \ll 1$), the low energy effective theory contains QCD but is hard to analyze using gauge-gravity duality \cite{m9711, gkp9802, w9802, agmoo}, while at strong coupling ($\lambda_4 \gg 1$) it can be effectively analyzed this way; however, the theory is no longer QCD but is instead a related theory known as holographic QCD, with similar qualitative features.  At strong coupling, and in the probe approximation $N_\f\ll N_c$, one can safely consider the flavor branes in the near-horizon geometry of ``color" $\D4$-branes.  The analysis of the flavor DBI action shows that the energetically-favored configuration of the flavor branes is one where they smoothly join at some radial point in the background geometry to form a U-shaped configuration where at the tip of the configuration the asymptotic $U(N_\f)\times U(N_\f)$ chiral symmetry is broken down to a single $U(N_\f)$.  By analyzing the near-horizon geometry of the color branes, one can also show that the model exhibits confinement and a mass gap for glueballs \cite{w9803, bisy9803}. 

One can separate the scale of chiral symmetry breaking from that of confinement by allowing the flavor branes to be separated by an asymptotic distance of $\ell_0 < \pi R_0$, instead of being placed at the antipodal points of the $x^4$-circle as in the Sakai-Sugimoto model \cite{ahjk0604}.  One can completely turn off confinement by taking the $R_0 \to \infty$ limit where the field theory dual is holographic NJL (which reduces to a nonlocal version of the NJL model at weak coupling).  At strong coupling, chiral symmetry is broken via a smooth fusion of the flavor branes, providing a holographic model of $\XSB$ without confinement.

These models can also be considered at finite temperature and chemical potential, where they exhibit similar behavior to QCD at finite temperature and chemical potential.  The finite-temperature analysis of the Sakai-Sugimoto model \cite{asy0604} and its non-compact version \cite{ps0604} amounts to putting the probe flavor branes in the near-horizon geometry of $N_c$ non-extremal $\D4$-branes.  The DBI action of the flavor branes indicates that at low temperatures (compared to $\ell_0^{-1}$) the energetically-favored solution is that of smoothly-connected $\D8$- and $\overline{\D8}$-branes which, like its zero-temperature counterpart, is a realization of chiral symmetry breaking.  At high enough temperatures, on the other hand, the preferred configuration is that of disjoint $\D8$- and $\overline{\D8}$-branes; hence, chiral symmetry is restored.  

A disadvantage of the Sakai-Sugimoto models is that, despite being models of spontaneous $\XSB$, the order parameter for such a breaking is absent.  In these models there is no mode in the bulk geometry from which one can extract holographically the order parameter for chiral symmetry breaking.  Also, related to this issue, one cannot write an explicit mass term for the localized fermions because there is no direction transverse to both the color and the flavor branes along which to stretch an open string.  So far, there are three proposals for how to modify the models in order to be able to compute the $\XSB$ order parameter and incorporate a bare fermion mass:
\begin{description}

\item{1)}
\underline{Open string tachyon}: Ref.  \cite{ahjk0604} argued that one should include the dynamics of an open string stretched between the flavor branes, whose scalar mode (tachyon) transforms as the bifundamental of $U(N_\f)\times U(N_\f)$ and has the right quantum numbers to be holographically dual to the fermion mass and condensate.  Refs \cite{ckp0702, bss0708, dn0708} incorporated this open string mode into the system using the tachyon-DBI action \cite{g0411} and showed that the tachyon has a normalizable and a non-normalizable mode, identified with the order parameter for $\XSB$ and the fermion mass, respectively.  
\item{2)} 
\underline{Open Wilson line}:  In holographic QCD and NJL, one can make a gauge-invariant operator out of the left-handed and right-handed fermions, $\psi_{iL}$ and $\psi^j_{R}$, which are localized at different points in the $x^4$-direction by including an open Wilson line between them.  The vev of this open Wilson line (OWL) operator is an order parameter for $\XSB$, argued in \cite{ak0803} to be holographically dual to a Euclidean string worldsheet bounded by the U-shaped flavor brane configuration.
\item{3)} 
\underline{Technicolor $\D4$-branes}: Ref.  \cite{hhly0803} added a technicolor sector to the original set-up by introducing ``technicolor" $\D4$-branes parallel to the color $\D4$-branes.  The flavor branes form a U-shaped configuration in the near-horizon geometry of the technicolor $\D4$-branes, and hence chiral symmetry is spontaneously broken in this sector.  A string stretched between the technicolor and the color $\D4$-branes is a gauge boson which can then mediate the chiral symmetry breaking of the technicolor sector to the QCD (or NJL) sector.  As a result, mass can be generated for the fermions of the QCD sector through a four-Fermi interaction.  To calculate the fermion mass, one has to calculate the 
area of a Euclidean worldsheet bounded by the $\D4$, $\D8$ and $\overline{\D8}$-branes.  
\end{description}

In this paper, we focus on the second of these proposals and use it to compute the $\XSB$ order parameter as a function of temperature and constant background electric and magnetic field.  In section 2, we briefly review holographic NJL, along with Aharony and Kutasov's proposal for computing the $\XSB$ order parameter $\langle\OWL\rangle$. In section 3, we investigate the effect of finite temperature on $\langle\OWL\rangle$.   We find the surprising result that  $\langle\OWL\rangle$ increases with temperature as one approaches the chiral phase transition temperature from below, even though the dynamically generated fermion mass decreases.  Both vanish above the transition temperature, as is expected in the chiral symmetry restoration ($\XSR$) phase.   We explore the possibility that this unexpected behavior of $\langle\OWL\rangle$ in the $\XSB$ phase is merely the reflection of the open Wilson line inside the operator rather than the fermion insertions at the endpoints but find evidence that this is not the case. In fact, we find that there are other models, such as the holographic Gross-Neveu and compact Sakai-Sugimoto models, for which  $\langle\OWL\rangle$ decreases with temperature or is independent of temperature, respectively. This suggests that  the behavior of  $\langle\OWL\rangle$ also strongly depends on the background in which the OWL operator is being studied. 

In section 4, we find that $\langle\OWL\rangle$ decreases (increases) with a constant background electric (magnetic) field, regardless of whether or not the temperature is turned on. At high enough  electric 
field (in some appropriate units), we show that  $\langle\OWL\rangle$ vanishes. Such behavior is also seen for  the usual chiral order parameter as a function of background electric or magnetic field in the standard NJL model, and is consistent with the inhibition of the chiral transition by a background electric field, as well as the catalysis of $\XSB$ by a background magnetic field.

In section 5, we demonstrate the existence of a family of generalized OWLs which correspond to the string worldsheet curving in additional directions. These different $\XSB$ order parameters in holographic NJL may correspond to open Wilson line operators with additional insertions in the five-dimensional gauge theory description.  By analyzing the global structure of their holographic dual worldsheets, we  show that, as expected,  the vevs of these  chirally-charged  generalized OWL operators vanish in the $\XSR$ phase of the model. In section 6, we consider more general OWL operators for which the fermions are located at two different points in spacetime, and show that their vevs also vanish in the $\XSR$ phase.

\section{Review of OWLs in holographic NJL}

In this section, we review the computation of the $\XSB$ order parameter in holographic NJL as proposed in \cite{ak0803}.  We start the section by first presenting a short review of the holographic NJL model \cite{ahjk0604}.  For simplicity, we set $N_\f=1$ in our discussion as the generalization to $N_\f \ll N_c$ is straightforward. 

\subsection{Holographic NJL}

Consider the non-compact Sakai-Sugimoto brane setup described in the introduction,
where the flavor D8- and $\overline{\D8}$-branes are separated by a coordinate distance $\ell_0$ in the non-compact $x^4$ direction along which the $N_c$ D4-branes are extended.  In this set-up, the left and right fermion modes, $\psi_L$ and $\psi_R$, are localized at D8-D4 and $\overline{\D8}$-D4 intersections and so interact via $\D4$-brane gauge fields (and scalars) whose strength is controlled by an effective dimensionless coupling $\lambda_{\rm eff}=\lambda_5/\ell_0$.  

At weak coupling, $\lambda_{\rm eff} \ll 1$, and in the regime where $g_s N_c$ is small so that stringy effects can be neglected, integrating out the $\D4$-brane gauge fields (and ignoring the interaction due to the exchange of $\D4$-brane scalars) generates an effective action for the fermions with a non-local interaction term  \cite{ahjk0604}
\bea\label{intnlnjl}
S_{\rm int} \sim g^2_5 \int d^4x d^4y G(x-y, \ell_0)\Big[\psi^{\dagger}_L(x)\cdot\psi_R(y)\Big]\Big[\psi^{\dagger}_R(y)\cdot\psi_L(x)\Big], 
\eea
where $G(x-y, \ell_0)$ is a $4+1$-dimensional scalar propagator and the dot in the parentheses denotes contraction over color indices. Thus, the theory of the fermions is a non-local version of the NJL model with a natural cut-off of $\Lambda=\ell_s^{-1}$. At weak coupling, the analysis of the gap equation shows \cite{ahjk0604} that $\psi^{\dagger}_L(x) \psi_R(y)$ develops a condensate breaking the chiral symmetry even at arbitrarily small values of $\lambda_{\rm eff}$. This is different from the standard NJL model where the condensate cannot form below a critical coupling. 

For $\lambda_{\rm eff} \sim 1$, the single gluon exchange approximation that leads to (\ref{intnlnjl}) is not valid.  At strong coupling ($\lambda_{\rm eff} \gg 1$), a better description of the system is by gauge-gravity duality. In the probe approximation, one considers the flavor branes in the near-horizon geometry created by $N_{\rm c}$ extremal $\D4$-branes given by the metric
\bea\label{Dq metric}
ds^2=\left(\frac{u}{R}\right)^{3/2}\Big(dt^2 + d{\vec x} ^2
\Big) + \left(\frac{u}{R}\right)^{-{3/2}}
\Big(du^2+ u^2 d{\Omega^2_{4}}\Big),
\eea
where $d\Omega^2_4$ is the metric of a unit 4-sphere. The characteristic length parameter $R$ is given by
\bea\label{radius}
R^{3}= \pi g_s  N_c l_s^{3}=\frac{1}{4\pi}
g_{5}^2 N_c l_s^{2},
\eea
where $g_s$ is the string coupling.  Also, the dilaton $\phi$ and the 4-form RR-flux $F_{(4)}$ are given by
\bea\label{dilatonq-form}
e^\phi=g_s \left(\frac{u}{R}\right)^{3/4},
\qquad F_{(4)} = \frac{2\pi N_c}{V_{4}}\ \epsilon_{(4)},
\eea
with $V_4$ and $\epsilon_{(4)}$ being the volume and the volume form of the unit $4$-sphere, respectively.

The dynamics of a $\D8$-brane (and a $\overline {\D8}$-brane) is determined by its DBI plus Chern-Simons action. Solving for the equations of motion for the gauge field, one can safely set the gauge field equal to zero and just work with the DBI part of the action
\bea\label{DBIaction}
S_{\rm{DBI}}=-\mu_8 \int d^{9}\sigma~e^{-\phi} \sqrt{-\hbox{det}(g_{ab})},
\eea
where $\mu_8$ is a constant and $g_{ab}=G_{MN}\del_a x^M \del_b x^N$ is
the induced metric on the worldvolume of the $\D8$-brane, with  $G_{MN}$ being the metric for the near horizon geometry of the color $D4$-branes. Since we would like to determine the shape of the flavor branes as a function of the radial coordinate $u$, we choose an embedding where the flavor branes form a curve $u=u(x^4)$ in the $(u, x^4)$-plane, extend in $\mathbb{R}^{3,1} \times S^4$ and are subject to the boundary condition 
\bea\label{Dbranebc1}
u(\pm \frac{\ell_0}{2})=u_{\Lambda} \to\infty.
\eea
The DBI action (\ref{DBIaction}) then reads
\bea\label{Dactionemb1}
S_{\rm{DBI}}= -C\int d^{3+1}x~dx^4 ~ u^{4} \sqrt{1+\Big(\frac{u}{R}\Big)^{-3}{ {u}^{'}}^2},
\eea
where $C= \mu_8 V_4/g_s$ and the primes here denote the derivative with respect to $x^4$. 

The first integral of the equation of motion for $u(x^4)$ is
\bea\label{braneeom}
\frac{u^{4} }{\sqrt {1+\Big(\frac{u}{R}\Big)^{-3}{u^{'}}^2}}=u_0^{4},
\eea
where $u_0$ parametrizes the solutions. The simplest solution compatible with 
the boundary condition (\ref{Dbranebc1}) is $x^4=\pm{\ell_0}/2$ which represents disjoint $\D8$- and $\overline {\D8}$-branes.  Note that this ``trivial" solution is obtained by setting $u_0=0$ in (\ref{braneeom}).  For $u_0\neq 0$, solving for $u^{'}$ yields
\bea\label{uprime}
{u^{'}}^2=\frac{1}{{u_0}^8} \, \Big(\frac{u}{R}\Big)^{3} ~\Big(u^8 -{u_0}^8\Big),
\eea
which shows that $u$ has a turning point at $u_0$.  This class of curved flavor brane solutions labeled by $u_0$ are interpreted as representing the breaking of chiral symmetry: at large $u$  the $U(N_{\f})\times U(N_{\f})$  symmetry is manifest whereas at $u=u_0$, where the branes join, there is just one  $U(N_{\f})$ factor.  For a fixed $\ell_0$, the curved solution is more energetically favored over the disjoint solution indicating that holographic NJL at strong coupling has a vacuum with $\XSB$.  

Integrating (\ref{uprime}) results in 
\bea\label{sol}
\int_0^{\ell_0/2} dx^4=R^{3/2}u_{0}^4 I_3 \qquad &\Rightarrow& \qquad
\ell_0=\frac{R^{3/2}}{4\sqrt{u_{0}}} B \Big( \frac {9}{16}, \frac {1}{2} \Big), 
\eea
where we defined the convenient set of integrals
\bea
I_n= \int_{u_0}^{\infty} \frac{du}{u^{n/2} \sqrt {u^8 -{u_0}^8}}=\frac{1}{8} u_0^{-\frac{1}{2} (n+6)}B\Big(\frac{n+6}{16}, \frac{1}{2}\Big), \qquad n>0.
\eea
Equation (\ref{sol}) implies that for each $\ell_0$ there exists a unique solution representing holographically a $\XSB$ phase of the dual gauge theory.  It also indicates that, the larger the asymptotic separation, the lower the $\XSB$ scale $u_0$.  Although equation (\ref{sol}) by itself does not put bounds on $\ell_0$, other considerations do so.  As $u_0\rightarrow 0$, one enters a regime of high curvature for which the supergravity approximation is no longer valid, putting an upper bound on  $\ell_0$.  Also, in order to ensure the stability (or metastability) of the flavor branes at large $u$, one has to require $\ell_0 \gg \ell_s$ which gives a lower bound on $ \ell_0$.

\subsection{OWLs as order parameters for $\XSB$}

Since $\psi_{L}$ and $\psi_{R}$ are localized at different points on $x^4$, a Wilson line must be inserted in the standard $\langle\psi^{\dagger j}_L \psi_{iR}\rangle$ $\XSB$ order parameter in order to render it gauge invariant.  Thus, we define the OWL operator  
\bea\label{owl}
{\OWL}^j_i(x^{\mu})=\psi^{\dagger j}_L(x^\mu, x^4=-\ell _0 / 2){\cal P}\ \hbox{exp}\Big[ 
\int_{-\ell _0 / 2}^{\ell _0 / 2}(i A_4 + \Phi)dx^4 \Big] \psi_{iR}(x^\mu, x^4=\ell _0 / 2),
\eea
where $A_4$ is the component of the gauge field in the $x^4$-direction, and $\Phi$ is a scalar.  Note that when $\psi_{iL}$ and $\psi^j_{R}$ are weakly-coupled to $A_4$ and $\Phi$, the operator defined in (\ref{owl}) reduces to the usual order parameter for chiral symmetry breaking in the NJL model. 

One can make many more non-local operators of the above type by choosing different contours for the Wilson line, or by inserting other operators along the contour.  These operators are interesting in their own right, and we will discuss some of them in later sections.  For now, we focus on the simplest such operator, given by (\ref{owl}). 

Analogous to holographic Wilson loops, it was proposed in \cite{ak0803} that at strong coupling the operator (\ref{owl}) is holographically dual to a Euclidean worldsheet bounded by the contour on which the operator is defined at $u=u_{\Lambda}$ and the flavor branes, and localized in the rest of the directions. Furthermore, the one-point function of $\OWL^j_i (x^{\mu})$ is given by 
\bea\label{AKprescription}
\langle \OWL^j_i \rangle\simeq \delta^j_i e^{-S_{\rm F}},
\eea
where $S_{\rm F}$ is the regularized action of a Euclidean fundamental string whose 
worldsheet was described above. The leading contribution to the string action is given 
by the  Nambu-Goto action
\bea\label{ng}
S_{\rm F}=\frac{1}{2\pi \alpha^{'}} \int d\sigma^1 d\sigma^2 \sqrt {\hbox{det} h_{\alpha \beta}}, \qquad \hbox{with} \qquad
h_{\alpha\beta}=G_{MN}\del_{\alpha} x^M \del_{\beta} x^N.
\eea
One can check that the worldsheet described above is a solution to the equations of motion arising from the Nambu-Goto action.

To calculate the regularized area (action) of this worldsheet we first choose the gauge $x^4=\sigma^1$ and $ u=\sigma^2$, and substitute this back into (\ref{ng}) to obtain
\bea\label{ws}
S_{\rm F}=\frac{1}{2\pi \alpha^{'}} \int dx^4 [u_\Lambda-u(x^4)],
\eea
where $u_\Lambda$ is the cutoff. By changing the measure of the integral to $du$, we obtain
\bea\label{ngd}
S_{\rm F}&=&\frac{\ell_0}{2\pi \alpha^{'}}u_\Lambda - \frac{1}{\pi \alpha^{'}} R^{3/2} u_{0}^4 I_1
\nonumber\\
&=&\frac{\ell_0}{2\pi \alpha^{'}}u_\Lambda-\frac{1}{8\pi \alpha^{'}}R^{3/2}\sqrt{u_{0}} B\Big( \frac {7}{16},\frac {1}{2} \Big).
\eea
For a finite result, the linear divergent piece in (\ref{ngd}) can either be absorbed within $\langle \OWL^j_i \rangle$ or subtracted away by a Legendre transform of the action (\ref{ngd}) with respect to $u$ \cite{dgo9904}. Using (\ref{radius}) and (\ref{sol}) together with $\lambda _5= (2\pi)^2 g_s N_c \ell_s$, one arrives at
\bea
S_{\rm F}=-c \lambda_{\rm eff},
\eea
where 
\bea\label{defc}
c= \frac{1}{128\pi^2}B\Big( \frac {7}{16},\frac {1}{2} \Big) B\Big( \frac {9}{16},\frac {1}{2} \Big)\approx 0.008.
\eea
Thus, (\ref{AKprescription}) yields
\bea\label{owll}
\langle \OWL^j_i \rangle\simeq\delta^j_i e^{c \lambda_{\rm eff} }.
\eea
The above result represents the leading contribution in $\lambda_{\rm eff}$ to $\langle \OWL^j_i \rangle$.  The next-to-leading contributions come from two sources: the coupling to the dilaton in the worldsheet action \cite{ft85}, and the fluctuation determinant around the saddlepoint. For the worldsheet in our discussion, these effects have been calculated in \cite{mms0807}. In this paper, we only consider the leading contribution in $\lambda_{\rm eff}$ to $\langle \OWL^j_i \rangle$.   

Note that the holographic computation leading to (\ref{owll}) is valid for $\lambda_{\rm eff} \gg 1$, which implies that $\langle \OWL^j_i \rangle$ is exponentially large. The exponential behavior of $\langle \OWL^j_i \rangle$ is unexpected from our experience with QCD, or the NJL model where the order parameter is related to the dynamically generated fermion mass. The unexpected behavior of $\langle \OWL^j_i \rangle$ has been attributed in \cite{ak0803} to the existence of the open Wilson line in the definition of the operator in (\ref{owl}), not the fermion bilinears. In what follows, we will see more surprises in the behavior of $\langle \OWL^j_i \rangle$ when we consider the model at different external conditions. 

\section{OWLs in thermal holographic NJL} 

\subsection{Holographic NJL at finite temperature}

To consider the holographic setup at finite temperature, one studies the dynamics of the flavor $\D8$- and $\overline {\D8}$-branes in the near-horizon geometry of a stack of non-extremal $\D4$-branes with the metric
\bea\label{bhmetric}
ds^2=\Big(\frac{u}{R}\Big) ^{3/2} \Big(-f(u)dt^2+d{\vec x}^2 \Big)+ \Big( \frac{u}{R}\Big)^{-{3/2}} \Big(\frac{du^2}{f(u)} +u^2 d\Omega_4^2\Big),\qquad f(u)=1-\frac{u_T^3}{u^3}, 
\eea
where $u_T$ is the horizon radius, which is related to the inverse temperature $\beta$ of the black brane by 
\bea\label{betadef}
u_T=\frac{16}{9}\frac{\pi^2 R^3 }{\beta^2}.
\eea
The dilaton and the 4-form RR-flux are the same as in the previous section.

\begin{figure}[h]
   \epsfxsize=6.0in \centerline{\epsffile{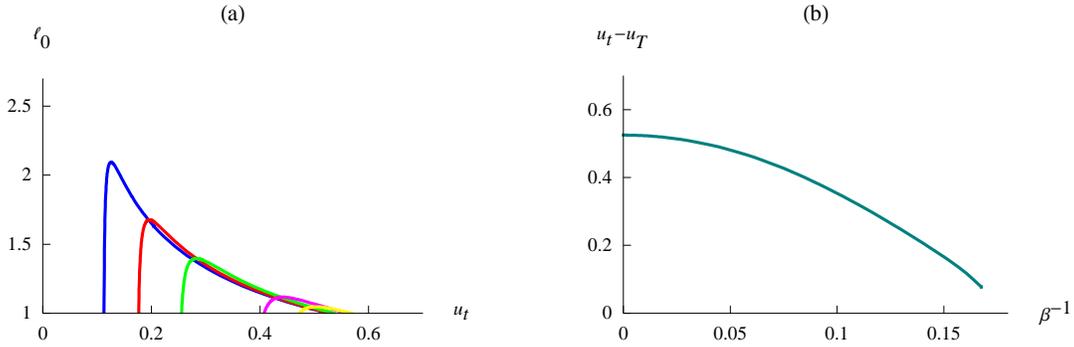}}
   \caption[FIG. \arabic{figure}.]{\footnotesize{ Plot (a) shows $\ell_0$ versus $u_t$ for various temperatures: $\beta^{-1}= 0.08$ (blue), $ 0.1$ (red), $0.12$ (green), $ 0.15$ (violet) and $ 0.16$ (yellow). For convenience, we set $R=1$ and cut the plots by the line $\ell_0=1$ in order to better show the values of $u_t$ for which $\ell_0=1$.  For $\beta^{-1}\gtrsim 0.167$, there exists no U-shaped flavor branes  with an asymptotic separation of $\ell_0=1$. Note that the behavior shown in the plots stays qualitatively the same for all $ \ell_0$'s. Plot (b) shows that $u_t-u_T$ decreases with temperature. The plot is  for $\ell_0=1$ but the qualitative behavior is  the same for other values of $\ell_0$ as well. As can be seen, the plot stops at $\beta^{-1}\approx0.167$ (beyond which there are no U-shaped solutions) but before reaching this temperature the system undergoes a $\XSR$ phase transition at $\beta^{-1}_\XSB\approx 0.15$. }}
\label{owl-one}
\end{figure}

In order to determine the vacuum, we use the same embedding and boundary condition as in the previous section. Hence, the DBI action for a $\D8$-brane in this geometry is given by
\bea
S_{\rm{DBI}}= -C\int d^{3+1}x~dx^4~ u^{4} \sqrt{f(u)+\Big(\frac{u}{R}\Big)^{-3}{u^{'}}^2}.
\eea
The equation of motion gives
\bea
{u^{'}}^2=\frac{1}{{u_0}^8} \Big(\frac{u}{R}\Big)^{3} f(u)[u^8 f(u)-{u_0}^8], 
\eea
which upon integration gives
\bea\label{L0}
\frac{\ell_0}{2}=R^{3/2} u_0^4 \int_{u_t}^{\infty} \frac{du} {u^{3/2} \sqrt{f(u)[u^8f(u)-u_0^8]}},
\eea
where the turning point $u_t$ is given by the largest real root of the equation
\be\label{tp}
u_t^8-u_T^3 u_t^5-u_0^8=0.
\ee
The trivial case of $u_0=0$ represents parallel flavor branes descending all the way down to the horizon $u=u_T$. This is interpreted as a phase of holographic NJL in which chiral symmetry is unbroken.  A given non-vanishing value of $u_0$ represents  a U-shaped flavor brane solution.  However, for a given value of $\ell_0$ at a particular temperature the analysis of (\ref{L0}) and (\ref{tp}) shows that there are two different values of  $u_t$.  Thus, there are two branches of solutions for which the flavor branes are smoothly joined at some turning point $u_t$. Figure \ref{owl-one} (a) shows plots of $\ell_0$ versus $u_t$ for different temperatures where, for convenience, we have fixed the asymptotic separation of the flavor branes to  $\ell_0=1$. We have also set $R=1$. We will refer to the solution with larger (smaller) turning point $u_t$ as the short (long) solution. For $\beta^{-1}\gtrsim 0.167$, there exists no U-shaped solution for which $\ell_0=1$.  For this temperature, Figure \ref{owl-one} (b) shows that $u_t-u_T$ does not vanish, indicating that the U-shaped solution ceases to exist before falling into the horizon.

Both short and long solutions are realizations of chiral symmetry breaking, though the constituent dynamical fermion mass $(u_t-u_T)/2\pi\alpha'$ is different for the two solutions.  Energy analysis shows that, below a critical temperature of $\beta^{-1}_{\chi \rm{SB}} \approx 0.15$ (in units of $\ell_0^{-1}$),  the short solution is favored over the solution with parallel branes (and also over the long solution); hence, chiral symmetry is broken. Beyond this critical temperature, the parallel solution is energetically favored so the vacuum represents a chiral symmetry-restored ($\XSR$) phase.  Note that from (\ref{tp}) it is easy to see that $u_t>u_0$ for non-vanishing $u_{T}$. For fixed $\ell_0$, the turning point increases as the horizon radius increases, but does so at a slower rate such that the two radii coalesce at high temperature. Figure \ref{owl-one} (b) shows that $u_t-u_T$  decreases as temperature increases (up to the phase transition temperature $\beta^{-1}_\XSB \approx 0.15$). Although the plot is for a specific value of $\ell_0=1$, its qualitative behavior stays the same for all $\ell_0$'s. In QCD or in the NJL model, if the dynamically generated fermion mass decreases as a function of temperature, one  concludes that the order parameter $\langle\psi^{\dagger j}_L \psi_{iR}\rangle$ should also decrease.  However, as we mentioned in the previous section, for the model under discussion there is no particular relationship between the dynamically generated fermion mass $(u_t-u_T)/2\pi\alpha'$ and the order parameter $\langle \OWL^j_i \rangle$. Thus, it is a matter of calculation to see how $\langle \OWL^j_i \rangle$ behaves as a function of temperature.

\subsection{OWLs in the $\chi$SB phase}

Let $\langle \OWL^j_i \rangle_\beta$ denote the order parameter at finite inverse temperature $\beta$.  In the chirally-broken phase of the model where the flavor branes smoothly join at a radial point above the horizon, one can show that there always exists a Euclidean worldsheet extended in the $(u, x^4)$-plane bounded by the flavor branes, and at a fixed point in the other directions. In order to calculate $\langle \OWL^j_i \rangle_\beta$, one has to find the regulated area of the Euclidean worldsheet in the black hole geometry (\ref{bhmetric}). We choose the same worldsheet embedding as before: $x^4=\sigma^1$ and $ u=\sigma^2$.
In this gauge, the action of the Euclidean worldsheet becomes
\bea\label{wsfinitet}
S_{\rm F}=\frac{1}{2\pi \alpha^{'}} \int dx^4 \int_{u(x^4)}^{u_\Lambda} \frac{du}{\sqrt{f(u)}}.
\eea

Although the above expression cannot be integrated in closed form, we can provide an analytic expression at low temperatures (compared to $\ell_0^{-1}$), $u_t\gg u_T$. From (\ref{tp}) one can easily see that $u_t\gg u_T$ implies that $u_0\gg u_T$, where $u_0$ is the minimum radius of the U-shaped flavor branes at zero temperature. Comparing (\ref{wsfinitet}) with (\ref{ws}) shows that $\langle \OWL^j_i \rangle_\beta$ should take the form 
\bea\label{wsgeneral} 
\langle \OWL^j_i \rangle_\beta= \delta^j_i {\rm{exp}}\Big[c \lambda_{\rm eff}\sum_{m=0} a_m 
\Big(\frac{\ell_0}{\beta}\Big)^m\Big],
\eea
where $a_m >0$ are to be determined by evaluating (\ref{wsfinitet}). Note that $a_0 =1$. The above expression can be further simplified. Since $f(u)=1-u_T^3/u^3$, the integral (\ref{wsfinitet}) indicates that only $a_{6k}$ in (\ref{wsgeneral}) are non-vanishing where $k \in \mathbb{Z}^{+}$. Therefore $\langle \OWL^j_i \rangle_\beta$ takes the general form
\bea\label{wsgeneralfinal} 
\langle \OWL^j_i \rangle_\beta= \delta^j_i {\rm{exp}}\Big[c\lambda_{\rm eff}\sum_{k=0} a_{6k} 
\Big(\frac{\ell_0}{\beta}\Big)^{6k}\Big].
\eea
In the following expressions, we will include the leading finite temperature corrections to the worldsheet action, namely we will keep terms up to the order $u_T^3/u_0^3$. Thus, (\ref{wsfinitet}) becomes
\bea\label{wsapp}
S_{\rm F}&\approx& \frac{1}{2\pi \alpha^{'}} \int dx^4 \int_{u(x^4)}^{u_\Lambda} du\ \Big( 1+\frac{u_T^3}{2u^3}\Big),\nonumber\\
&=& \frac{\ell_0}{2\pi \alpha^{'}}u_\Lambda\Big( 1-\frac{u_T^3}{4u_\Lambda^3}\Big)-\frac{1}{\pi \alpha^{'}} R^{3/2} u_0^4 \int_{u_t}^{\infty} \Big( 1+\frac{u_T^3}{4u^3}\Big)\frac{du} {u^{1/2} \sqrt{u^8-u_T^3 u^5-u_0^8}},   
\eea
where
\bea
u_t\approx u_0\Big( 1+\frac{u_T^3}{8u_0^3}\Big).
\eea
The following change of variable
\bea
u=v\Big( 1+\frac{u_T^3}{8v^3}\Big),
\eea
enables one to write the integral expression in (\ref{wsapp}) as
\bea\label{wsur}
S_{\rm F} = \frac{\ell_0}{2\pi \alpha^{'}}u_\Lambda\Big( 1-\frac{u_T^3}{4u_\Lambda^3}\Big)- \frac{1}{\pi \alpha^{'}} R^{3/2} u_0^4 \Big(I_1-\frac{u_T^3}{16}I_7\Big).
\eea

\paragraph {\bf Digression on the Legendre-transformed string action.} One may wonder whether a Legendre transform of the action along the lines of \cite{dgo9904} will take care of the dependence of the worldsheet action on $u_{\Lambda}$. Although performing the Legendre transform with respect to $u$ gets rid of the linear divergent term in (\ref{wsur}) \cite{cg0810}, it cannot cancel the subleading terms in $u_{\Lambda}$, and in fact introduces new counter-terms to the action. Define
\bea\label{defaction}
S_{\rm F} \to S_{\rm F}-\int dx^4 \sqrt{\hbox{det}(h_{\alpha \beta})}~\pi_u u, 
\eea  
where the integral is over the boundary of the worldsheet, $h_{\alpha \beta}$ is the boundary metric and  
\bea
\pi_u=\frac{1}{2\pi \alpha^{'}} G_{uu} n^{\mu} \del_{\mu}u,
\eea
is the momentum pulled back to the boundary and $n^{\mu}$ is a normal outgoing vector to the boundary. Calculating $\hbox{det}(h_{\alpha \beta})$ and $\pi_u$ at $u=u_{\Lambda}$, we obtain 
\bea\label{momen}
\hbox{det}(h_{\alpha \beta})=\Big(\frac{u_{\Lambda}}{R}\Big)^{3/4}, \qquad \pi_u= \frac{1}{2\pi \alpha^{'}} \frac{1}{f(u_{\Lambda})} \Big(\frac{u_{\Lambda}}{R}\Big)^{-{3/4}}.
\eea
Substituting (\ref{momen}) into (\ref{defaction}) and expanding $f(u_{\Lambda})$ for small temperatures, we arrive at
\bea
S_{\rm F} \to \frac{\ell_0}{2\pi \alpha^{'}}u_\Lambda\Big( 1-\frac{u_T^3}{4u_\Lambda^3}\Big)- \frac{1}{\pi \alpha^{'}} R^{3/2} u_0^4 \Big(I_1-\frac{u_T^3}{16}I_7\Big)-\frac{\ell_0}{2\pi \alpha^{'}}u_\Lambda\Big( 1+\frac{u_T^3}{u_\Lambda^3}\Big),
\eea
showing that the Legendre transform of the action removes the linear divergent piece (but leaves dependence on the cutoff in non-divergent terms).

\vspace{18pt}

Nevertheless, in the $u_\Lambda \to \infty$ limit, one arrives at an unambiguous result for the action
\bea\label{wsr}
S_{\rm F} = - \frac{1}{\pi \alpha^{'}} R^{3/2} u_0^4 \Big(I_1-\frac{u_T^3}{16}I_7\Big).
\eea
At low temperatures, $\ell_0$ can also be approximated as
\bea
\frac{\ell_0}{2} &\approx& R^{3/2} u_0^4 \Big(I_3+\frac{u_T^3}{16}I_9\Big).
\eea
Expressing $u_0$ in terms of $\ell_0$ and substituting the result in (\ref{wsr}), one obtains 
\bea
S_{\rm F} = -c\lambda_{\rm eff} \Big[1+ a_T\Big(\frac{\ell_0}{\beta}\Big)^{6}\Big], 
\eea
where $c$ has already been defined in (\ref{defc}) and 
\bea\label{defk}
a_T=4\left[\frac{8}{3}\frac{\pi}{B\Big( \frac {9}{16},\frac {1}{2} \Big)}\right]^6 
\left[\frac{B\Big( \frac {15}{16},\frac {1}{2} \Big)}{B\Big( \frac {9}{16},\frac {1}{2} \Big)}-
\frac{B\Big( \frac {13}{16},\frac {1}{2} \Big)}{B\Big( \frac {7}{16},\frac {1}{2} \Big)}\right]\approx 136.2.
\eea
Thus, we obtain 
\bea
\frac{\langle \OWL^j_i \rangle_\beta}{\langle \OWL^j_i \rangle}=1+1.09\lambda_{\rm eff} \Big(\frac{\ell_0}{\beta}\Big)^{6}.
\eea

\begin{figure}[h]
   \epsfxsize=2.5in \centerline{\epsffile{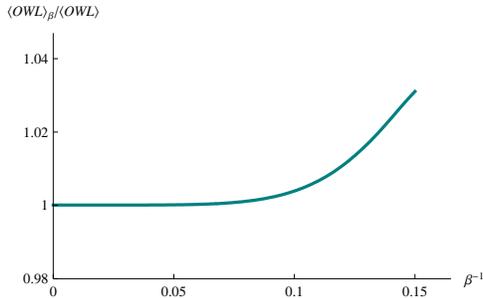}}
   \caption[FIG. \arabic{figure}.]{\footnotesize{$\langle \OWL^j_i \rangle_\beta/\langle \OWL^j_i \rangle$ increases with temperature in the chirally-broken phase of the model, and the chiral restoration transition is first order. The plot is for  $\ell_0=1$ and, for convenience, we set  $R=1$, $2\pi\alpha^{\prime}=1$ and $u_{\Lambda}=10$. The plot stays qualitatively the same for larger values of $u_{\Lambda}$.}}
\label{owl-two}
\end{figure}

In order to demonstrate that this behavior persists as one increases the temperature, we numerically solve for $\langle \OWL^j_i \rangle_\beta$. For convenience, we set $R=1$, $\ell_0=1$ and $2\pi\alpha^{\prime}=1$, resulting in $\lambda_5= 8\pi^2$. In order for the IR sensitivity of the OWL not to be smeared out by integrating too far in the UV region, we choose the cutoff $u_{\Lambda}=10$.  As we vary the temperature, we use a shooting algorithm to find the values of $u_t$ which correspond to keeping $\ell_0=1$. Figure \ref{owl-two}  not only shows  that $\langle \OWL^j_i \rangle_\beta$ increases with temperature but also that the transition to the chirally-symmetric phase of the model is first order.  One can check that behavior of the plot stays qualitatively the same for other values of $\ell_0$ as well. We have cut the plot at $\beta^{-1}_{\chi \rm{SB}}\approx 0.15$ where the chiral restoration phase transition happens \cite{ps0604, eln0803} but, in principle, one can continue the plot beyond $\beta^{-1}_{\chi \rm{SB}}$ up to $\beta^{-1} \lesssim 0.167$ which is the critical temperature (for a fixed $\ell_0=1$) beyond which no U-shaped solutions exist.  

Our analytic and numerical results in this section both show that, for holographic NJL at strong coupling and in the the $\XSB$ phase, $\langle \OWL^j_i \rangle_\beta$ increases monotonically with temperature. This behavior is quite different from what one observes in QCD and in the standard NJL model. Indeed, in the standard NJL model at finite temperature, the analysis of the gap equation for the order parameter $\langle\psi^{\dagger j}_L \psi_{iR}\rangle$ shows that it decreases with temperature. Plotting the effective potential for the order parameter $V_{\rm eff} (\langle\psi^{\dagger j}_L \psi_{iR}\rangle)$ versus $\langle\psi^{\dagger j}_L \psi_{iR}\rangle$ for various temperatures, one observes that the absolute minimum of  the plot moves towards smaller values of $\langle\psi^{\dagger j}_L \psi_{iR}\rangle$ as the temperature is increased (see for example Figure 1 in \cite{rd87}). At a critical temperature, the minimum is at $\langle\psi^{\dagger j}_L \psi_{iR}\rangle=0$, for which the model goes to the chirally-restored phase. 

\subsection{Contour-cancelled $\chi$SB order parameters}

Although $\langle \OWL^j_i \rangle$ goes over to the usual order parameter $\langle\psi^{\dagger j}_L \psi_{iR}\rangle$ at weak coupling for the holographic NJL model, its unusual temperature dependence raises the question of whether there exist other gauge-invariant chirally-charged operators similar to (\ref{owl}) whose vev shows the behavior expected of an order parameter in the NJL model.  In fact, there are infinitely many OWL operators which can act as chiral symmetry-breaking order parameters.  For instance, the shape of the Wilson line contour can be changed and other local operators can be inserted at points along its length.  (Some examples of these modified order parameters will be examined in detail in sections 5 and 6, below.)  

Here, let us try to use this freedom to define other order parameters to test one possible explanation (already mentioned in \cite{ak0803}) for the unexpected behavior of $\langle \OWL^j_i\rangle$ both at zero and at finite temperature: namely, that the value of its vev is dominated by the open Wilson line contour inside the operator rather than the fermion insertions at the end points.   
A way of testing this is to consider ratios of operators with different Wilson line contours.  In particular, an appropriate ratio could (approximately) cancel the contour dependence of the OWL, leading to a $\XSB$ order parameter whose vev is dominated by the fermion insertions.  So, does such a contour-cancelled order parameter decrease with increasing temperature?  We will now argue that it does not.  This suggests that the unusual behavior of the OWL vev is not due to its being dominated by the Wilson line as opposed to the fermion insertions, and that the temperature-dependence of the OWL order parameter reflects a physical property of the $\XSB$ transition in holographic NJL, and is not just an artifact of a poor choice of order parameter.

\begin{figure}[h]
   \epsfxsize=3.0in \centerline{\epsffile{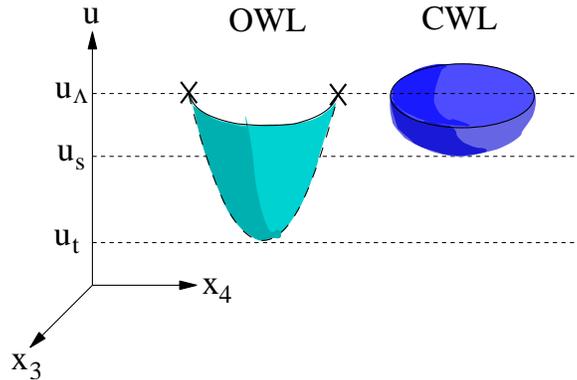}}
   \caption[FIG. \arabic{figure}.]{\footnotesize{The shaded areas represent string worldsheets ending on a semi-circular open Wilson line in the $x_3$-$x_4$ plane $u=u_\Lambda$ (green) and on a circular Wilson loop of the same radius in the same plane (blue).   For the open Wilson line, the X's correspond to the endpoint sources and the worldsheet's dashed boundary lies along the D8-brane, which is not shown.  $u_t$ is the altitude of the D8 brane turning point, while $u_s$ is the altitude of the turning point of the circular Wilson loop's string worldsheet.}}
\label{owl-alter}
\end{figure}

One way to cancel the contour dependence is to divide the OWL operator by the square root of the vev of the closed Wilson loop (CWL) along a contour that traverses the open contour in one direction then returns along the same contour to its starting point.  However this clearly does not work because the CWL vev has no temperature dependence as the area enclosed by the contour vanishes.  A better choice is to open up the CWL contour to reduce such cancellations.  For example, a particularly simple and symmetrical choice is to choose the CWL to be a circle in the $x^3$-$x^4$ plane of
radius $\ell_0/2$, and to choose the OWL contour to be the semicircle of the same radius between the fermion insertions.  In computing the saddlepoint contribution, we consider the corresponding worldsheet for the CWL to have boundary on the circle at $u=u_\Lambda$, while the worldsheet for the OWL has  boundaries on the semicircle at $u=u_\Lambda$ and on the D8-branes.  These worldsheets are roughly depicted in Figure \ref{owl-alter}. 
Then we can construct our contour-cancelled order parameter as
\be
O^j_i := {\OWL^j_i\over \sqrt{\langle \CWL \rangle}},
\ee
so that
\be
\langle O^j_i \rangle\simeq\delta^j_i e^{-(S_{\rm F_{open}}-\frac{1}{2} S_{\rm F_{closed}})},
\ee
where $S_{\rm F_{open}}$ is the action of the open string worldsheet with the semi-circular contour and D8-branes as its boundary, and $S_{\rm F_{closed}}$ is the action of the worldsheet with the circular Wilson loop as its boundary.  

Note that this kind of contour-cancelled order parameter has the advantage of  automatically being UV regulated, since the contributions from the string worldsheets near $u=u_\Lambda$ cancel.  A disadvantage of this type of order parameter, however, is that having the contour bend in some spacetime direction ($x^3$ in the figure) necessarily breaks space-time rotational invariance, which was preserved by 
Aharony and Kutasov's OWL order parameter proposal.

It is difficult to analytically evaluate the dependence of this contour-cancelled order parameter on the temperature.  For example, the equation governing the string worldsheet with boundary on the CWL is derived from the Nambu-Goto action.  Choosing polar coordinates $(r,\phi)$ in the $x^3$-$x^4$ plane and assuming that the worldsheet is $\phi$-independent so that $u=u(r)$, the Nambu-Goto action is proportional to
$$
S_{\rm NG} \propto \int_0^{r_0}\!\!dr\, r \sqrt{u^3 + {u^3 {u'}^2\over u^3 - u_T^3}},
$$
where we have rescaled $r \to R^{3/2}\ r$ to remove extraneous factors, so $r_0 = (\ell_0/2) R^{-3/2}$.  This gives the equation of motion
$$
0 = 2ru (u^3-u_T^3) u'' + 2 u {u'}^3 - 3r (2u^3-u_T^3) {u'}^2 + 2u(u^3-u_T^3) u' 
- 3 r (u^3-u_T^3)^2,
$$
with boundary conditions $u'=0$ at $r=0$ and $u=u_\Lambda$ at $r=r_0$.  This has to be solved numerically.

The general result is that the CWL string worldsheet has a much smaller area than the OWL worldsheet, and so that $\sqrt{\langle\CWL\rangle}\gg\langle\OWL\rangle$.  This is reflected in the fact that the altitude $u_s$ of the CWL string worldsheet is much higher than the turning point $u_t$ of the D8-brane.  This is simple to understand qualitatively:  the fact that the D8-brane is extended in more dimensions than the string means that its area scales as a higher power of $u$, and so decreases its area by descending to lower values of $u$.  Thus, the
the turning point $u_t$ of the D8-brane is much more sensitive to changes of the temperature (and, therefore  to the location of the horizon $u_T$) than is the turning point $u_s$ of the CWL string.  For example, for $\ell_0=0.5$ (in units of $R$), $u_t$ varies approximately between $2.10$ and $2.25$ as the temperature increases from 0 to the $\XSR$ transition, while $u_s$ varies only from $38.772$ to $38.774$ over the same temperatures, which is only about  1/75 of the first variation.

Since the area of the worldsheets is essentially proportional to $\ell_0(u_\Lambda-u_s)$ or $\ell_0(u_\Lambda-u_t)$ for the CWL or OWL worldsheets, respectively, it follows that the temperature dependence of $u_t$ completely dominates that of $u_s$.  Thus, the exponential increase of the order parameter with temperature is hardly any different between the original $\langle\OWL\rangle$ order parameter, and the contour-corrected one, $\langle O\rangle$.

\subsection{$\chi$SB order parameter in holographic Gross-Neveu}

Although $\langle \OWL^j_i\rangle$ does not show the expected behavior of an order parameter in the $\XSB$ phase of holographic NJL, it may well be the case that it shows the expected behavior (that is to say, it decreases with temperature) for other intersecting brane models  of chiral symmetry breaking and restoration. To put it another way, $\langle \OWL^j_i\rangle$ may not only strongly depend on the choice of a contour,  but also on the background in which it is embedded.

For instance, consider a $\D2$-$\D8$-$\overline{\D8}$ system where $N_{\f}$ flavor $\D8$- and $\overline{\D8}$-branes intersect $N_c$ color $\D2$-branes at two (1+1)-dimensional intersections, and are separated by a distance $ \ell_0$, say, in the $y$-direction. The massless degrees of freedom at the intersections are left-handed and right-handed Weyl fermions which interact via the exchange of gauge fields (and scalars) of the color theory whose strength is given by an effective coupling $\lambda_{\rm eff}= \lambda_3 \ell_0$ . At small effective coupling, after integrating out the gauge fields, one arrives at a generalized Gross-Neveu (GN) model with a non-local four-Fermi interaction \cite{gxz0605} \footnote{There is another holographic realization of the GN model based on intersecting $\D4$-$\D6$-$\overline{\D6}$-branes which gives rise to a non-local four-Fermi interaction  \cite{ahk0608}.}. At strong coupling, one studies the dynamics of the flavor branes in the near-horizon geometry of extremal $\D2$-branes. This theory, which we call holographic GN, like holographic NJL, breaks chiral symmetry.  It also shows a $\XSR$ transition at a finite temperature. 

To study the model at finite temperature and large coupling and in the probe approximation,  one considers the flavor branes in the near-horizon geometry of non-extremal $\D2$-branes, which is described by the metric 
\bea\label{d2action}
ds^2=\Big(\frac{u}{R_3}\Big) ^{5/2} \Big(-h(u)dt^2+d{x}^2 +d{y}^2\Big)+ \Big( \frac{u}{R_3}\Big)^{-{5/2}} \Big(\frac{du^2}{h(u)} +u^2 d\Omega_6^2\Big), 
\eea
where 
\bea
R_3^5=2\pi^2\lambda_3l_s^6, \qquad h(u)=1-\frac{u_T^5}{u^5}.
\eea
The temperature of the black brane is related to $u_T$  by 
\bea\label{d2temp}
\frac{1}{\beta}=\frac{5}{4\pi}\Big(\frac{u_T}{R_3}\Big)^{5/2}\frac{1}{u_T}.
\eea
The dilaton and the 4-form RR-flux of the solution take the forms
\bea
e^\phi=g_s \left(\frac{u}{R_3}\right)^{-5/4},
\qquad F_{(2)} = \frac{2\pi N_c}{V_{6}}\ \epsilon_{(6)},
\eea
where $V_{6}$ and $\epsilon_{(6)}$ are the volume and the volume form of the unit $6$-sphere, respectively. 

For a $\D8$-brane which wraps the 6-sphere and forms a curve $u=u(y)$, the DBI action reads
\bea
S_{\rm{DBI}}= -\frac{\mu_8}{g_s} {\rm Vol}(S^6)R_3^{5/2}\int d^{1+1}x~dy~ u^{7/2} \sqrt{h(u)+\Big(\frac{u}{R_3}\Big)^{-5}{u^{'}}^2}.
\eea
The equation of motion gives
\bea\label{d2braneeom}
\frac{u^{7/2}h(u)}{\sqrt {h(u)+\Big(\frac{u}{R}\Big)^{-5}{u^{'}}^2}}=u_0^{7/2},
\eea
where, as usual, $u_0=0$ is a parallel brane solution which is not energetically favored at low temperatures. For $u_0\neq 0$, we have (at small temperatures)
\bea\label{d2L0}
\frac{\ell_0}{2}&=&R^{5/2} u_0^{7/2} \int_{u_t}^{\infty} \frac{du} {u^{5/2} \sqrt{h(u)[u^7-u^2u_T^5-u_0^7]}},\nonumber\\
&\approx&R^{5/2} u_0^{7/2}\Big(K_5-\frac{2}{7}u_T^5K_{15}\Big),
\eea
where 
\bea
K_n= \int_{u_0}^{\infty} \frac{du}{u^{n/2} \sqrt {u^7 -{u_0}^7}}=\frac{1}{7} u_0^{-\frac{1}{2} (n+5)}B\Big(\frac{n+5}{14}, \frac{1}{2}\Big), \qquad n>0.
\eea

The action of a Euclidean worldsheet bounded by the flavor branes,  stretched in the $(u, y)$-plane and point-like in other directions is given by 
\bea\label{d2wsfinitet}
S_{\rm F}&=&\frac{1}{2\pi \alpha^{'}} \int dx^4 \int_{u(x^4)}^{u_\Lambda} \frac{du}{\sqrt{h(u)}},\nonumber\\
&\approx&\frac{\ell_0}{2\pi \alpha^{'}}u_\Lambda\Big( 1-\frac{u_T^5}{8u_\Lambda^3}\Big)- \frac{1}{\pi \alpha^{'}} R^{5/2} u_0^{7/2} \Big(K_3-\frac{23}{56}u_T^5K_{13}\Big),
\eea
where in the second line we have only kept the leading term in the small temperature approximation. Eliminating $u_0$ between (\ref{d2L0}) and (\ref{d2wsfinitet}), dropping the $u_\Lambda$-dependent terms in (\ref{d2wsfinitet}), and keeping the leading  temperature-dependent term, we get
\bea
S_{\rm F} = -b(\lambda_{\rm eff} )^{1/3}\Big[1-b_T\Big(\frac{\ell_0}{\beta}\Big)^{10/3}\Big], 
\eea
where $b$ and $b_T$ are given by 
\bea\label{defk2}
b&=&\frac{1}{7}\left[\frac{28}{5 B\Big( \frac {5}{7},\frac {1}{2} \Big)}\right]^{1/3} B\Big( \frac {4}{7},\frac {1}{2} \Big)\approx 0.63, \nonumber\\
b_T&=&\frac{1}{56}\left[\frac{14}{\pi}\frac{\pi}{B\Big( \frac {5}{7},\frac {1}{2} \Big)}\right]^{10/3} 
\left[23\frac{B\Big( \frac {9}{7},\frac {1}{2} \Big)}{B\Big( \frac {4}{7},\frac {1}{2} \Big)}-8
\frac{B\Big( \frac {10}{7},\frac {1}{2} \Big)}{B\Big( \frac {5}{7},\frac {1}{2} \Big)}\right]\approx 10.5.
\eea
So, we have
\bea
\frac{\langle \OWL^j_i \rangle_\beta}{\langle \OWL^j_i \rangle}=1-6.61\lambda_{\rm eff} ^{1/3}\Big(\frac{\ell_0}{\beta}\Big)^{6}, 
\eea
showing that the $\langle \OWL^j_i \rangle_\beta$ decreases with temperature in the small temperature approximation. 

\begin{figure}[h]
   \epsfxsize=6.0 in \centerline{\epsffile{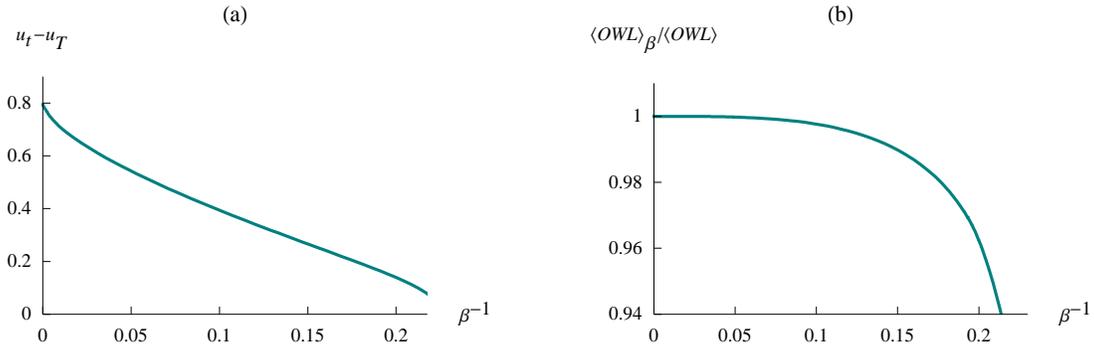}}
   \caption[FIG. \arabic{figure}.]{\footnotesize{For U-shaped branes for which $\ell_0=1$, we set $R_3=1$, $2\pi\alpha^{\prime}=1$ and $u_{\Lambda}=10$, and  numerically plotted (a) $u_t-u_T$,  and (b) $\langle \OWL^j_i \rangle_\beta/\langle \OWL^j_i \rangle$ versus temperature $\beta^{-1}$.  The plots show that, in the $\XSB$ phase of holographic GN, both $u_t-u_T$ and  $\langle \OWL^j_i \rangle_\beta$ decrease with temperature , and the transition is first order.}}
\label{owl-three}
\end{figure}

This behavior persists for higher temperatures, as shown by numerical calculations.  For example, in Figure \ref{owl-three} we show the results of such calculations where we have set $R_3=1$, $2\pi\alpha^{\prime}=1$ and $u_{\Lambda}=10$, and plotted $\langle \OWL^j_i \rangle_\beta/\langle \OWL^j_i \rangle$, as well as  $u_t-u_T$, versus temperature $\beta^{-1}$ for U-shaped branes with an asymptotic separation of $\ell_0=1$. With our choice of parameters, $\lambda_{\rm eff}= 4\pi$, indicating that holographic GN is in the strong-coupling regime.  Note that, although in the plots of Figure \ref{owl-three} we considered temperatures up to  $\beta^{-1}\approx 0.22$ beyond which there is no U-shaped solution with $\ell_0=1$, the system undergoes a $\XSR$ phase transition at $\beta^{-1}_\XSB\approx 0.2$.  Figure \ref{owl-three} (b) shows that for temperatures up to  $\beta^{-1}_\XSB$, $\langle \OWL^j_i \rangle_\beta$ decreases with temperature. Also, since $\langle \OWL^j_i \rangle_\beta$ does not vanish at $\beta^{-1}_\XSB$, the $\XSR$ phase transition is first order. The qualitative behavior of the plots presented in Figure \ref{owl-three} is independent of the particular choice of the cut-off  $u_{\Lambda}$ we have made,  and also stays the same for all allowed values of $\ell_0$.

\subsection{$\chi$SB order parameter in the (compact) Sakai-Sugimoto model}

As another example, consider the Sakai-Sugimoto model, where $x^4$ is now a circle with radius $R_0$ and the flavor branes  are located at the antipodal points of the circle, $\ell_0=\pi R_0$. At finite temperature, there are two competing background geometries from which one determines the deconfinement temperature $\beta_c^{-1}$. The deconfinement temperature is obtained when the two geometries are equally energetically favorable. At low temperatures when the model is in the confined phase, the background geometry is the same as the zero temperature geometry \cite{ss0412} but with Euclidean time periodically identified. Also, in this phase, the flavor $\D 8$- and $\overline {\D8}$-branes smoothly join at the tip of the background geometry, thereby realizing chiral symmetry breaking. At high temperatures, the model is in the deconfined phase and the background is identical to (\ref{bhmetric}) except that  the $x^4$-direction is a circle. In this phase, the preferred configuration is that of disjoint flavor branes. Hence, chiral symmetry is restored. One can easily show that, in the confined phase where chiral symmetry is also broken, $\langle \OWL^j_i \rangle_\beta$ does not change with temperature and equals its value at zero temperature \cite{ak0803} 
\bea
\langle \OWL^j_i \rangle_\beta=\langle \OWL^j_i \rangle \simeq \delta^j_i e^{\lambda_5/18\pi R_0}, \qquad \forall~ \beta^{-1} <\beta_c^{-1}.
\eea
This is the expected behavior of the $\XSB$ order parameter $\langle\psi^{\dagger j}_L \psi_{iR}\rangle$ in the confined phase of finite temperature QCD  \cite{ng83}. Also,  $\langle \OWL^j_i \rangle_\beta$ vanishes in the deconfined phase (see section 3.6 for more details). 

For the generalized Sakai-Sugimoto model 
where $\ell_0<\pi R$, there also exists an intermediate phase \cite{asy0604} for which the theory is 
deconfined while chiral symmetry is broken. For this phase, an analysis similar to what we did in this section shows that $\langle \OWL^j_i \rangle_\beta$ increases with temperature up to the $\XSR$ temperature. 

\subsection{OWLs in the $\chi$SR phase}

In the high-temperature phase (temperatures large compared to $\ell_0^{-1}$) of the model, the flavor branes are parallel. There exists a Euclidean worldsheet, bounded by the parallel $\D 8$- and $\overline {\D8}$-branes, which penetrates into the horizon.  In order to better understand the behavior of this worldsheet, especially inside the horizon, we use Kruskal coordinates.  This will be useful later in our discussion of modified $\XSB$ order parameters in section 5.  

To reduce clutter, we scale out the coordinates of the metric by replacing 
\bea
u\to u_T~u, \qquad x^\mu \to R^{3/2} u_T^{-1/2}~x^\mu,
\eea
so that the background metric (\ref{bhmetric}) takes the form 
\bea\label{nodimbhmetric2}
\frac{ds^2}{\gamma^2} = u^{3/2} \Big[-f(u)dt^2 + d{\vec x}^2\Big]
+ u^{-3/2} \Big[f(u)^{-1} du^2 + u^2 d\Omega_4^2\Big],
\eea
where $f(u) = 1-u^{-3}$, and $\gamma^2= {4\pi\over3\beta} R^3$. Note that in (\ref{nodimbhmetric2}), $t, \vec x$ and $u$ are all dimensionless.

\begin{figure}[h]
   \epsfxsize=2.5in \centerline{\epsffile{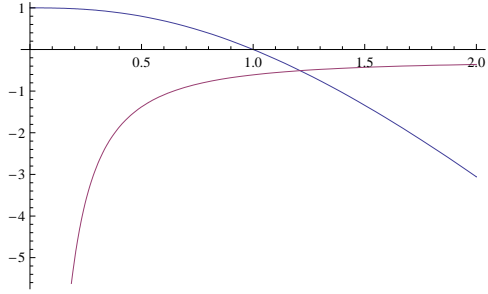}}
   \caption[FIG. \arabic{figure}.]{\footnotesize{$vw(u)$ (upper, blue curve) and $G_{vw}(u)$ (lower, red curve).}}
\label{owl-krusk1}
\end{figure}

First, define the tortoise coordinate $r$ by
\be\label{tort}
{dr\over du} = {u^{3/2}\over u^3-1}.
\ee
We choose phases in the solution to (\ref{tort}) so that near $u=1$ we have $e^{3r}\approx u-1$.  Define the infalling and outgoing Kruskal coordinates by
\bea\label{kruscoor}
v := +e^{+3(t+r)/2},
\qquad
w := -e^{-3(t-r)/2},
\eea
so that the metric becomes
\be\label{actionkrus}
\frac{ds^2}{\gamma^2}= -{4\over9}{u^3-1\over u^{3/2}}e^{-3r} dvdw + u^{3/2} d{\vec x}^2+ u^{1/2} d\Omega_4^2.
\ee
Note that $vw=-e^{3r}$, so $r$ and $u$ are functions only of the combination $vw$.  Figure \ref{owl-krusk1} shows $vw$ 
and $G_{vw}$ as functions of $u$.  Note that the singularity ($u=0$) is at $vw=1$ and the horizon $(u=1$) is at $vw=0$.  

\begin{figure}[h]
   \epsfxsize=2in \centerline{\epsffile{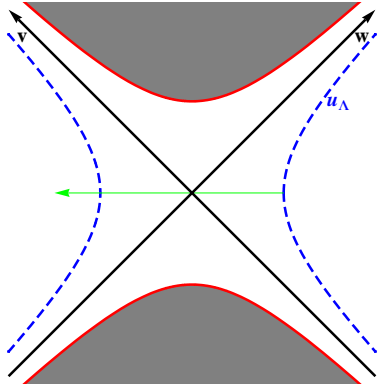}}
   \caption[FIG. \arabic{figure}.]{\footnotesize{The near-horizon geometry of black $\D4$-branes in Kruskal coordinates. The blue dashed lines represent cutoffs and the red lines are the singularities.}}
\label{owl-krusk2}
\end{figure}

Figure \ref{owl-krusk2} shows the black hole geometry in Kruskal coordinates.  The red lines are the singularities, the blue dashed lines are the UV cutoff(s), and the $v$ and $w$ axes are the horizons.    In Kruskal coordinates the extended black hole geometry is apparent, including the second asymptotic region and the past ``white hole" ($w<0$).

The straight string worldsheet descending from the OWL between two straight D8-branes is extended along $u$ and $x^4$ at fixed values of the other (Schwarzschild-like) coordinates.  In particular, it is at a fixed value of $t$, say $t=t_0$.  This corresponds to the line $w = -e^{-3t_0} v$ in Kruskal coordinates, and is shown as the green line in Figure \ref{owl-krusk2}.  This line passes into the second asymptotic region where it has nowhere to end, as there is no $\OWL$ insertion at $u=u_\L$ there for it to end on.  Therefore, in the $\XSR$ phase of the model there exists no finite-area worldsheet \cite{mms0807} to be holographically dual to $\OWL^j_i (x^{\mu})$; hence, $\langle \OWL^j_i \rangle =0$. 

We can also examine the OWL worldsheet in Euclidean signature, where the Euclidean time $t_E$ is periodically identified with the inverse temperature: $t_E \simeq t_E + \beta$. The period $\beta$ is given by (\ref{betadef}), so that the horizon becomes a regular point with local polar coordinates approximately given by $(\rho,\phi)\approx(\,\sqrt{\beta(u-u_T)/\pi} \,, \,2\pi t_E/\beta\,)$ near the horizon, $u= u_T$.  The worldsheet descends from the OWL at constant $t_E$, say $t_E=0$, until it reaches $\rho=0$. At this point, it passes through the origin and emerges in the antipodal direction, $t_E=\beta/2$.  The worldsheet continues to arbitrarily large $u$ in this direction, since there is no OWL insertion at $t_E=\beta/2$ for it to end on.  Thus, it again gives zero saddlepoint contribution to $\langle \OWL\rangle$, consistent with being in the $\XSR$ phase.

\section{Holographic NJL OWLs in background electromagnetic fields}

In this section, we determine how  $\langle \OWL^j_i \rangle$ responds to turning on a background electric or magnetic field in holographic NJL.  We turn on a $U(1)$ background electromagnetic field by gauging the $U(1)_V$ part of the $U(N_\f)_V$ symmetry of the model.  We adopt the convention of \cite{bll0802} (see also \cite{jk0803,ksz0803}) where the fermions have charge one under this $U(1)$ field; namely, $A=N^{-1}_\f \hbox{Tr} {\cal A}$ where $A \in U(1)$ and ${\cal A} \in U(N_{\f})$. We only consider the case $N_\f=1$ here.

\subsection{Background electric field}

To add a background electric field to the holographic NJL model in, for example, the $x^1$-direction, we turn on the $A_1$ component of the gauge field on the flavor branes and choose the ansatz
\bea\label{aaz}
2\pi\alpha' A_1= -Et. 
\eea
For a $\D8$-brane forming a curve ${x^4}=x^4(u)$ in the background of black $\D4$-branes (\ref{bhmetric}), and with the gauge field turned on as in (\ref{aaz}), the DBI action reads
\bea\label{eaction}
S_{\rm{DBI}}&=& -\mu_8 \int d\sigma^9 e^{-\phi} \sqrt{-\hbox{det}(g_{ab}+2\pi\a^{'}F_{ab})},\nonumber\\
&=&-C\int d^{3+1}x~du~u^4 \sqrt{\Big[\Big(\frac{u}{R}\Big)^{-3}+ f(u) \Big(\frac{dx^4}{du}\Big)^2\Big] \Big[ 1-\frac{1}{f(u)}\Big(\frac{u}{R}\Big)^{-3} E^2\Big] },
\eea
where $F_{ab}=F_{\mu \nu} {\partial}_{a}x^{\mu}{\partial}_{b} x^{\nu}$ is the induced field strength on the 
$\D 8$-branes, $C=\mu_8 V_4/g_s$, and $E$ is the dimensionless electric field. 
There is also a Wess-Zumino term 
\bea
S_{WZ}=\frac{N_c}{24\pi^2}\int \omega_5 (A)= 
\frac{N_c}{24\pi^2}\int A\wedge F \wedge F, 
\eea
which vanishes in our case. The equation of motion for the brane embedding $x^4 (u)$ is
\bea\label{eomze}
u^4 f(u)\left(\frac{1-\frac{1}{f(u)}\Big(\frac{u}{R}\Big)^{-3} {E}^2}{\Big(\frac{u}{R}\Big)^{-3}+ f(u) \Big(\frac{dx^4}{du}\Big)^2}\right)^{1/2} \frac{dx^4}{du} =u_{*}^4.
\eea

In (\ref{eaction}) the first term under the square root is positive, independent of the embedding. However, the second term can be negative, rendering the embedding unphysical. Indeed, it is the analog of the critical electric field for  flat branes and is easy to understand. Consider an open string with end points on the flavor branes which is also assumed to lie along the $x^1$-direction. The tension of this string is equal to its energy per unit length. Assume that at $u=u_{\Lambda} \to \infty$ the string tension is $(2\pi\alpha')^{-1}$. Therefore, its effective tension at a radius $u$ equals $(2\pi\alpha')^{-1} \sqrt{f(u)}~(u/R)^{3/2}$. Now an electric field in the $x^1$-direction will cause the string endpoints to move away from one another. If the string is at $u< u_{\rm cr}$, the electric field will overcome the string tension and will pair create. This is where the semi-classical description of the flavor brane in terms of its DBI action breaks down. This critical radius $u_{\rm cr}$ is where the string tension cancels the electric field and is given by  
\bea\label{ucr}
u_{\rm cr}^3=u_T^3+R^3{E}^2,
\eea
reproducing the radius at which the second factor in (\ref{eaction}) vanishes.  Below this critical radius, no physical embedding of the flavor branes can exist. 

Since (\ref{ucr}) indicates that the horizon is always below the critical radius, the parallel embedding of the flavor branes (the chirally-symmetric phase) cannot exist. The holographic interpretation of this result is that turning on even a small electric field, ${E} \ll 1$, in holographic NJL will wash out its otherwise would-be chirally-symmetric phase. Of course this interpretation is solely based on the DBI analysis of the flavor branes. As the flavor branes approach $u_{\rm cr}$ the electric field will break open strings and destabilize the system. Taking the corrections into account, it is plausible that a parallel embedding of the flavor branes can survive, resulting in a restoration of chiral symmetry in the model. A U-shaped configuration, on the other hand, can exist as long as the minimum radius $u_t$ of the solution stays above $u_{\rm cr}$.  

One consequence of having this critical radius around is that there is a bound on the dynamical constituent mass of the fermions. There is a critical radius for an open string stretched from the tip of the flavor branes to the horizon of the geometry. The mass (in string units) of this string equals $u_t-u_T$. Since the flavor branes cannot descend below $u_{\rm cr}$, the mass of the string (the dynamical constituent mass of the fermion) cannot be less than $u_{\rm cr}-u_T$. Equivalently, since $u_t-u_T$ is related to the asymptotic distance $\ell_0$ between the flavor branes, one deduces that there is a maximum $\ell_{\rm max}$ beyond which there is no U-shaped flavor branes. 

To simplify the equation of motion while still capturing the essentials  of turning on a background electric field, we first consider the zero temperature limit of the equation, thus describing holographic NJL at zero temperature but with a finite electric field. 

\subsubsection*{Zero temperature} 

At zero temperature, $f(u)=1$, in which case the critical radius  becomes $\bar u_{\rm cr}^3=R^3 {E}^2$. The $u_{*}=0$ solution in (\ref{eomze}), which represents the parallel branes, is not a physical solution for the reasons explained above, at least in the DBI approximation employed throughout this paper.  Thus, we assume $u_{*}\neq 0$. Note that, in order to trust the calculations, we assume that both $u_t$ and $\bar u_{cr}$ stay away from the region of high curvature of the background geometry. Setting $f(u)=1$ in (\ref{eomze}) and solving for $dx^4/du$, we get
\bea\label{zerox}
\Big(\frac{dx^4}{du}\Big)^2= \Big(\frac{u}{R}\Big)^{-3} \frac{u_{*}^8}{u^8-u^5 \bar u_{\rm cr}^3-u_{*}^8}, \qquad \bar u_{\rm cr}^3=R^3 {E}^2.
\eea
When there is no electric field, the minimum radius for the U-shaped flavor branes (the turning point) is at $u_t=u_{*}$. Turning on a small $E$ field, one expects a small deviation in $u_t$ from $u_{*}$. Assuming $\frac{u_{*}}{\bar u_{\rm cr}}\ll 1$, we obtain 
\bea
u_t \approx u_{*} \left[ 1+\frac{1}{8} \Big(\frac{\bar u_{\rm cr}}{u_{*}}\Big)^{3} \right].
\eea

\begin{figure}[h]
   \epsfxsize=6.0in \centerline{\epsffile{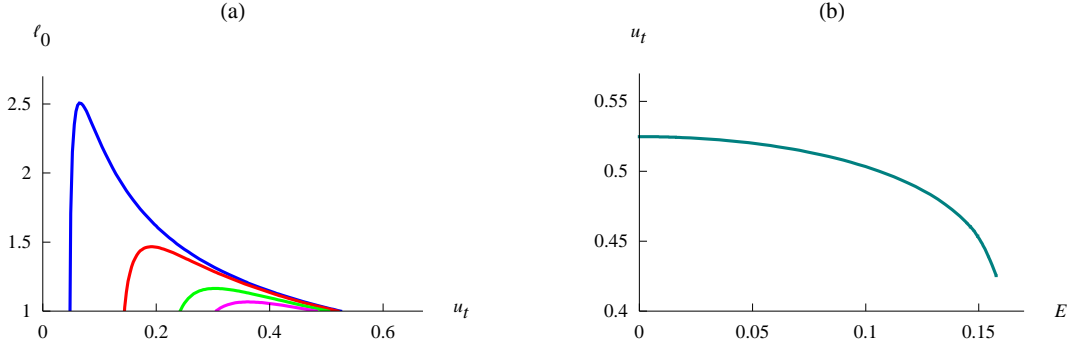}}
   \caption[FIG. \arabic{figure}.]{\footnotesize{(a) $\ell_0$ versus $u_t$ for different values of the electric field: $E= 0.01$ (blue), $ 0.05$ (red), $0.1$ (green) and $ 0.13$ (violet). We set $R=1$ and cut the plot by the  $\ell_0=1$ line, since we are interested in the number of solutions with an asymptotic separation of $\ell_0=1$. The plot shows that there are two branches of solutions approaching each other as $E$ approaches a maximum $E_{\rm max}\approx 0.158$, beyond which there is no solution with $\ell_0=1$. (b)  The turning point of the energetically-favored solution decreases with the electric field, reaching $u_{\rm min}\approx 0.43$.}}
\label{owl-four}
\end{figure}

To analyze the behavior of the solution near the critical radius, one has to take a different limit. As $u_{*}$ decreases to arbitrarily small but positive values (to exclude $u_{*}=0$ from the limit), the turning point approaches the critical radius. To leading order, we have
\bea
u_t \approx \bar u_{\rm cr} \left[ 1+ \frac{1}{3}\Big(\frac{u_{*}}{\bar u_{\rm cr}}\Big)^{8} \right] .
\eea
Figure  \ref{owl-four} (a) shows the number of U-shaped branes with a fixed asymptotic separation of $\ell_0=1$.  For the plot, we set $R=1$. There is a maximum electric field $E_{\rm max}\approx 0.158$, beyond which no U-shaped flavor branes with  $\ell_0=1$ can exist. For $E<E_{\rm max}$, there always exist two solutions (similar behavior is also observed for other values of $\ell_0$). Using the energy argument, one can show that the solution with larger $u_t$ is always energetically favored. Figure \ref{owl-four} (b) shows the plot of $u_t$ of the energetically-favored solution (with  $\ell_0=1$) versus $E$. Denoting the turning point (of the energetically-favored solution) corresponding to the maximum electric field by $u_{\rm min}$, we observe that $u_{\rm min} \approx 0.43$. Note that $u_{\rm min}> u_{\rm cr}({E_{\rm max}}) \approx 0.29$. The fact that  $u_{\rm min}> u_{\rm cr}$ indicates that the $E_{\rm max}$ of the boundary theory (holographic NJL) is not related to the maximum electric field on the flavor branes associated with the breaking of open strings on the branes at $u=u_{\rm cr}$.

Thus, as one increases the electric field, the energetically favored U-shaped brane goes further into the bulk until its turning point reaches $u_{\rm min}$. At the same time, the turning point of the less energetically favored solution increases with the electric field until it, too, reaches  $u_{\rm min}$. The picture emerging is that the two U-shaped solutions approach each other as electric field increases, and coalesce when $E=E_{\rm max}$.

Since the Euclidean worldsheet from which $\langle \OWL^j_i\rangle$ is calculated does not directly couple to the $A_1$ gauge field on the brane, one deduces from Figure \ref{owl-four} (b) that the (regulated) area of the worldsheet increases with the electric field. Hence,  $\langle \OWL^j_i\rangle$ decreases quadratically with the electric field, as long as the electric field is small. This is exactly the behavior observed in the standard NJL model for the dependence of the dynamically generated fermion mass on a constant background electric field \cite{kl89}. 

To calculate the leading-order dependence of $\langle \OWL^j_i\rangle$ on the electric field, we first find the area of the Euclidean worldsheet. After a bit of algebra, one arrives at 
\bea\label{actionze}
S_{\rm F}&=& \frac{1}{2\pi \alpha^{'}} \int dx^4 \int_{u(x^4)}^{u_\Lambda} du, \nonumber\\
&=& \frac{\ell_0}{2\pi \alpha^{'}}u_{\Lambda}-\frac{1}{\pi \alpha^{'}} R^{3/2} u_{*}^4 \int_{u_t}^{\infty} \frac{du} {u^{1/2} \sqrt{u^8-u^5 \bar u_{\rm cr}^3-u_{*}^8}},\nonumber\\  
&\approx& \frac{\ell_0}{2\pi \alpha^{'}}u_\Lambda - \frac{1}{\pi \alpha^{'}} R^{3/2} u_{*}^4 \Big (I_1-\frac{5}{16}\bar u_{\rm cr}^3 I_7\Big).
\eea
In general, $S_{\rm F}$ admits the expansion
\bea
S_{\rm F}&=& \frac{\ell_0}{2\pi \alpha^{'}}u_\Lambda - \frac{1}{8\pi \alpha^{'}} R^{3/2} u_{*}^{1/2}\sum_{n=0} s_n \Big(\frac{\bar u_{\rm cr}}{u_{*}}\Big)^{3n}.
\eea
Next, we solve for $u_{*}$ from 
\bea
\frac{\ell_0}{2}&=& R^{3/2} u_{*}^4 \int_{u_t}^{\infty} \frac{du}{ u^{3/2}\sqrt{u^8 -u^5 \bar u_{cr}^3-u_{*}^8}},\nonumber\\
&\approx& R^{3/2} u_{*}^4 \Big (I_3-\frac{7}{16}\bar u_{\rm cr}^3 I_9\Big).
\eea
Substituting $u_{*}$ back into (\ref{actionze}) and removing the linear divergent term from the action, we arrive at
\bea\label{ace}
S_{\rm F}=-c\lambda_{\rm eff}\left[1-a_E\Big(\frac{\ell_o}{R}\Big)^6 E^2\right],
\eea
where
\bea\label{defae}
a_E=2^8\Big[ B\Big( \frac {13}{16},\frac {1}{2} \Big)\Big]^{-6}
\left[5\frac{B\Big(\frac {13}{16},\frac {1}{2} \Big)}{B\Big( \frac {7}{16},\frac {1}{2} \Big)}+7
\frac{B\Big( \frac {15}{16},\frac {1}{2} \Big)}{B\Big( \frac {9}{16},\frac {1}{2} \Big)}\right]\approx 3.6,
\eea
and $c$ is the same as in (\ref{defc}). Thus, $\langle \OWL^j_i\rangle$ reads 
\bea\label{owlE}
\frac{\langle \OWL^j_i \rangle_{E}}{\langle \OWL^j_i \rangle}= 1-0.029\lambda_{\rm eff}\Big(\frac{\ell_o}{R}\Big)^6 {E}^2,
\eea
where $\langle \OWL^j_i \rangle_{E}$ denotes the order parameter for holographic NJL with a background electric field turned on. This shows that, in the $\XSB$ phase of the model, $\langle \OWL^j_i \rangle_{E}$ \emph{decreases} with electric field, at least for a small electric field. 
\begin{figure}[h]
   \epsfxsize=2.5in \centerline{\epsffile{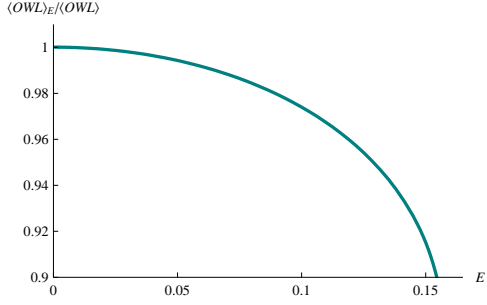}}
   \caption[FIG. \arabic{figure}.]{\footnotesize{At zero temperature, $\langle \OWL^j_i \rangle_{E}$ decreases with the electric field $E$. To obtain the plot, we set $\ell_0=1$,  $R=1$, $2\pi\alpha'=1$, and  $u_{\Lambda}=10$. When the electric field reaches its maximum value of  $E_{\rm max}\approx 0.158$,  the worldsheet reaches the radius $u_t=u_{\rm min}\simeq 0.43$, thereby minimizing  $\langle \OWL^j_i \rangle_{E}$ to $\langle \OWL^j_i \rangle_{E} \approx0.9~\langle \OWL^j_i \rangle$. }}
\label{owl-five}
\end{figure}

For arbitrary values of the  electric field up to $E_{\rm max}\approx 0.158$,  we have solved the equations numerically and a plot of $\langle \OWL^j_i \rangle_{E}/\langle \OWL^j_i \rangle$ versus $E$ is given in Figure \ref{owl-five}.  To obtain the plot, we kept  $\ell_0=1$, $R=1$ and $2\pi\alpha'=1$.  Also, in order to avoid smearing out the IR sensitivity of $\langle \OWL\rangle$ by integrating too far in the UV region, we chose the cutoff $u_{\Lambda}=10$. Again, the qualitative behavior of the plot is not sensitive to the particular choice of the parameters we have made.  The result obtained in Figure \ref{owl-five} qualitatively agrees with field theoretic computations in the standard NJL model (see for example \cite{kl89}), where turning on a constant background electric field has been shown to inhibit chiral symmetry breaking.  

\subsubsection*{Finite temperature} 

Having done the analysis for zero temperature, the computation at finite temperature is almost the same.  We first do the computations when both electric field and temperature are small: ${E}\ll 1$ and ${\ell_0}/\beta\ll 1$.  From (\ref{eomze}),
\bea\label{smalltx}
\Big(\frac{dx^4}{du}\Big)^2= \frac{1}{f(u)}\Big(\frac{u}{R}\Big)^{-3} \frac{{u_{*}}^8}{u^8-u^5 u_{\rm cr}^3-u_{*}^8}, \qquad u_{\rm cr}^3=u_T^3+ \bar u_{\rm cr}^3,
\eea
where $\bar u_{\rm cr}$ is given in (\ref{zerox}). To ensure the reality of the solution, the $u_{*}=0$ branch is excluded. Since the critical radius $u_{\rm cr}$ is always above the horizon, the turning point is the largest real root of $u^8-u^5 u_{\rm cr}^3-u_{*}^8=0$. To leading order in temperature and electric field, the turning point $u_t$ is 
\bea
u_t\simeq u_*\Big(1+\frac{u_{\rm cr}^3}{8u_{*}^3}\Big) .
\eea

Integrating (\ref{smalltx}) yields 
\bea\label{ellte}
\frac{\ell_0}{2}&=& R^{3/2} u_{*}^4 \int_{u_t}^{\infty} \frac{du}{ u^{3/2} \sqrt{f(u)}\sqrt{u^8 -u^5 u_{cr}^3-u_{*}^8}},\nonumber\\
&\approx& R^{3/2} u_{*}^4 \Big (I_3+ \frac{1}{16}u_T^3 I_9-\frac{7}{16}\bar u_{\rm cr}^3 I_9\Big).
\eea
Again, one can show that a Euclidean worldsheet extended in the $(x^4,u)$-plane, bounded by the flavor branes, and fixed in the other directions is a solution to the equation of motion obtained from the Nambu-Goto action. The action of this Euclidean worldsheet now becomes
\bea\label{actepele}
S_{\rm F}&=&\frac{1}{2\pi \alpha^{'}} \int dx^4 \int_{u(x^4)}^{u_\Lambda} \frac{du}{\sqrt{f(u)}},\nonumber\\
&\approx& \frac{\ell_0}{2\pi \alpha^{'}}u_\Lambda\Big( 1-\frac{u_T^3}{4u_\Lambda^3}\Big)-\frac{1}{\pi \alpha^{'}} R^{3/2} u_{*}^4 \int_{u_t}^{\infty}  \Big( 1+\frac{u_T^3}{4u^3}\Big)\frac{du}{u^{1/2}\sqrt{u^8-u_{cr}^3 u^5-u_{*}^8}},\nonumber\\  
&\approx& -\frac{1}{\pi \alpha^{'}} R^{3/2} u_{*}^4 (I_1-\frac{1}{16}u_T^3 I_7-\frac{5}{16}\bar u_{\rm cr}^3 I_7).
\eea
where in the last line above, we subtracted the linear divergent term and ignored the subleading terms in $u_\Lambda$.
Eliminating $u_{*}$ between (\ref{ellte}) and (\ref{actepele}), we get
\bea
S_{\rm F} = -c\lambda_{\rm eff} \Big[1+ a_T\Big(\frac{\ell_0}{\beta}\Big)^{6}-a_E\Big(\frac{\ell_o}{R}\Big)^6 {E}^2\Big],
\eea
where $c$, $a_T$ and $a_E$ are given in (\ref{defc}), (\ref{defk}) and (\ref{defae}), respectively. Thus, one obtains
\bea\label{owlET}
\frac{\langle \OWL^j_i \rangle_{E}^{\beta}}{\langle \OWL^j_i \rangle}= 1+1.09\lambda_{\rm eff} \Big(\frac{\ell_0}{\beta}\Big)^{6}-0.029 \lambda_{\rm eff}\Big(\frac{\ell_o}{R}\Big)^6 {{E}}^2,
\eea
Since the leading terms in $\beta^{-1}$ and $E$ do not mix, $\langle \OWL^j_i \rangle_{E}^{\beta} /\langle \OWL^j_i \rangle$ retains the same qualitative behavior that it had when either $\beta^{-1}$ or $E$ vanished.

\begin{figure}[h]
   \epsfxsize=6.0in \centerline{\epsffile{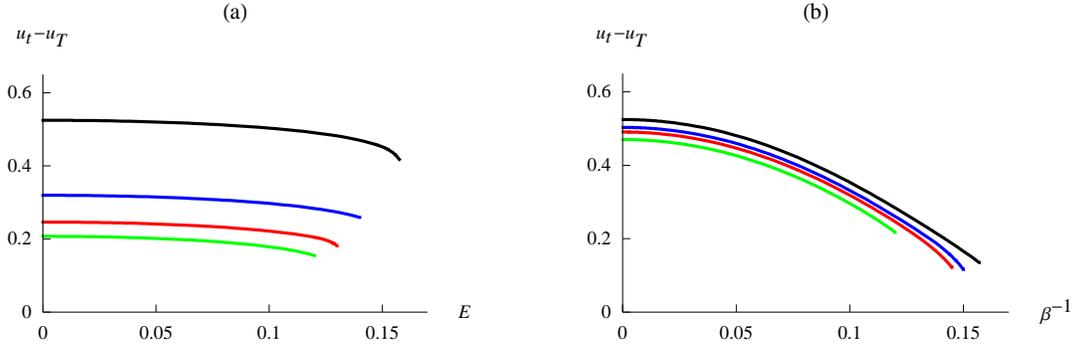}}
   \caption[FIG. \arabic{figure}.]{\footnotesize{ (a) $u_t-u_T$ versus $E$ for various temperatures: $\beta^{-1}= 0$ (black), $ 0.11$ (blue), $0.13$ (red) and $ 0.14$ (green). (b) $u_t-u_T$ versus $\beta^{-1}$ for different values of the electric field: $E=0$ (black), $ 0.1$ (blue), $0.12$ (red) and $ 0.14$ (green).  The plots in (a) and (b) are for U-shaped solutions with $\ell_0=1$ (and $R=1$)  but  the qualitative behavior is the same for all $ \ell_0$.}}
\label{elec-owl-six}
\end{figure}

We have also done the numerical calculations for arbitrary values of temperature and electric field, for an energetically favored U-shaped brane with a fixed asymptotic separation of $\ell_0=1$. Figure  \ref{elec-owl-six}(a) shows $u_t-u_T$ versus $E$ for fixed values of $\beta^{-1}$. This shows that the maximum electric field $E_{\rm max}$, beyond which the U-shaped solution (with $\ell_0=1$) cannot exist, decreases with temperature. Figure \ref{elec-owl-six}(b) shows $u_t-u_T$ versus $\beta^{-1}$ for fixed values of the electric field, which demonstrates that the maximum temperature beyond which our  U-shaped solution does not exist decreases with the electric field. 

\begin{figure}[h]
   \epsfxsize=5.5in \centerline{\epsffile{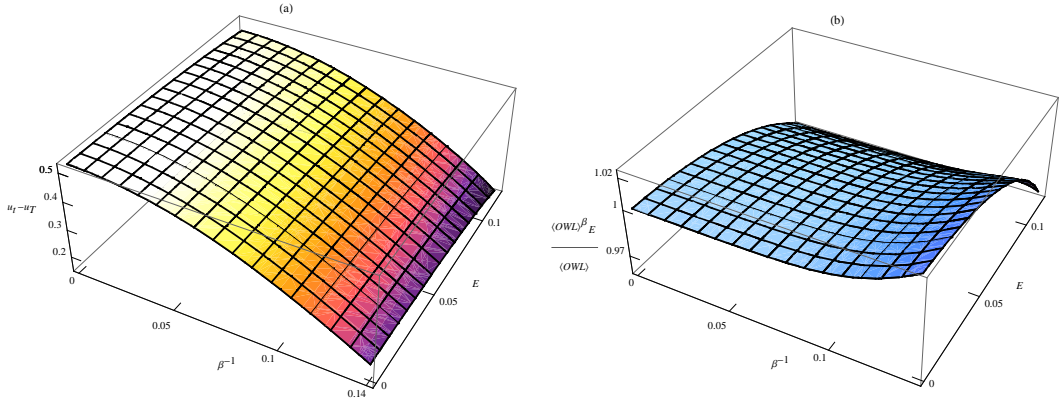}}
   \caption[FIG. \arabic{figure}.]{\footnotesize{ (a) $u_t-u_T$ as a function of the electric field $E$ and temperature $\beta^{-1}$. (b) $\langle \OWL^j_i \rangle_{E}^{\beta} /\langle \OWL^j_i \rangle$ as a function of $E$ and $\beta^{-1}$. These plots are for U-shaped solutions with $\ell_0=1$ (and $R=1$)  but the qualitative behavior is the same for all $ \ell_0$.}}
\label{elec-owl-seven}
\end{figure}

A three-dimensional plot of $u_t-u_T$ as a function of $\beta^{-1}$ and $E$ is given in Figure \ref{elec-owl-seven}(a), which shows that $u_t-u_T$ monotonically decreases with both $\beta^{-1}$ and $E$. $\langle \OWL^j_i \rangle_{E}^{\beta} /\langle \OWL^j_i \rangle$ as a function of $\beta^{-1}$ and $E$ is shown in Figure \ref{elec-owl-seven}(b). Note that $\langle \OWL^j_i \rangle_{E}^{\beta} /\langle \OWL^j_i \rangle$ monotonically decreases with $E$ but, unlike $u_t-u_T$, it monotonically increases with $\beta^{-1}$. In both directions, the transition to the chirally-restored phase is first order. 

We would like to emphasize that we wanted to know how $\langle \OWL^j_i \rangle$ responds to turning on a background electric field in holographic NJL, which is why we took the ansatz (\ref{aaz}). We could have generalized this ansatz by taking $2\pi\alpha'A_1(t,u)=Et+g(u)$ where, from the asymptotic behavior of $g(u)$, one reads off the current (in the $x^1$-direction) in the boundary theory. One then finds that, in the $\XSB$ phase, the energetically favored solution is the one with no current turned on. This is to say that the vacuum is in an insulating phase with broken chiral symmetry (see \cite{bll0802, jk0803, ksz0803} for more details). Thus, turning on a current will have no bearing on the computations we performed here for calculating $\langle \OWL^j_i \rangle$ in the $\XSB$ phase of the model. However, turning on a current, one finds that there exists a special value of the electric field (fixed in terms of the current) for which a parallel embedding of the flavor branes can be physical \cite{bll0802}. Hence, the high-temperature vacuum is a conductor with no chiral symmetry breaking. Since the string worldsheet from which we calculated 
$\langle \OWL^j_i \rangle$ does not couple to $A_1$ and is bounded by parallel branes, the general analysis in section 3.4 for $\langle \OWL^j_i \rangle$ in the high-temperature phase of the model goes through. That is, the string worldsheet penetrates the horizon and extends to the second asymptote of the background geometry, in which there is no $\OWL$ insertion to end on. Therefore, in this chirally-restored conducting phase of the model, $\langle \OWL^j_i \rangle$ also vanishes.

\subsection{Background magnetic field}

To turn on a constant background magnetic field in, for example, the $x^3$-direction of the boundary theory, we choose the following ansatz for the $U(1)$  gauge field on the $\D8$-brane
\bea
2\pi\alpha' A_2=H x^1,
\eea
where $H$ is dimensionless. The DBI action for the $\D8$-brane reads
\bea\label{maction}
S_{\rm{DBI}}=-C\int d^{3+1}x~du~u^4 \sqrt{\Big[\Big(\frac{u}{R}\Big)^{-3}+ f(u) \Big(\frac{dx^4}{du}\Big)^2\Big] \Big[ 1+\Big(\frac{u}{R}\Big)^{-3} {H}^2\Big] }.
\eea
The first integral of the resulting equation of motion for $x^4(u)$ is
\bea\label{meom}
u^4 f(u)\left(\frac{1+\Big(\frac{u}{R}\Big)^{-3} H^2}{\Big(\frac{u}{R}\Big)^{-3}+ f(u) \Big(\frac{dx^4}{du}\Big)^2}\right)^{1/2} \frac{dx^4}{du} =u_{*}^4.
\eea

Since both factors inside of the square root are positive, unlike the electric field, the magnetic field does not impose a condition for an embedding to be physical. This implies that both parallel and U-shaped embeddings exist. The $u_{*}=0$ branch in (\ref{meom}) represents the parallel embedding where holographic NJL is in the chirally-symmetric phase. Since  the $A_2$ gauge field does not couple to the Euclidean worldsheet (which is only extended in the $(x^4,u)$-plane), it neither changes the equation of motion nor the boundary conditions. Therefore, the worldsheet behaves as if there is no magnetic field. As we showed in the previous section, such a worldsheet dips inside the horizon and extends to the second asymptotic region of the black hole geometry. This implies that this worldsheet cannot be dual to a one point function of the $\OWL$ operator. Hence, in the chirally-symmetric phase of holographic NJL with a constant background magnetic field one must have $\langle \OWL^j_i \rangle=0$.

We will now analyze how a constant background magnetic field affects $\langle \OWL^j_i \rangle$ in the $\XSB$ phase.

\subsubsection*{Zero temperature}

\begin{figure}[h]
   \epsfxsize=6.0in \centerline{\epsffile{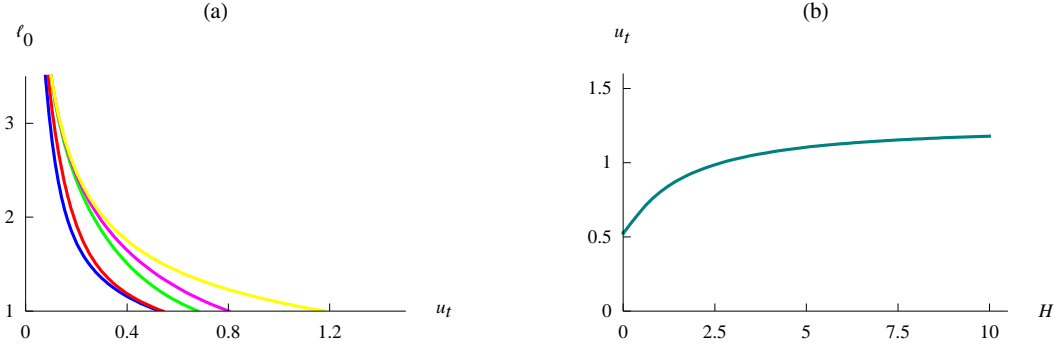}}
   \caption[FIG. \arabic{figure}.]{\footnotesize{(a) $\ell_0$ versus $u_t$ for different values of the magnetic field: $H= 0.05$ (blue), $ 0.1$ (red), $0.5$ (green), $1$ (violet) and $10$ (yellow). We set $R=1$ and cut the plot at the  $\ell_0=1$ line, in order to magnify the turning point of the solution with an asymptotic separation of $\ell_0=1$.  For each value of the magnetic field, there is only one U-shaped solution. (b) The turning point $u_t$ of the solution monotonically increases with the magnetic field $H$, approaching an asymptotic value at large $H$.}}
\label{mag-owl-eight}
\end{figure}

We turn off the temperature by setting $f(u)=1$ in (\ref{maction}) and (\ref{meom}). Solving for $dx^4/du$ gives
\bea\label{meomz}
\Big(\frac{dx^4}{du}\Big)^2= \Big(\frac{u}{R}\Big)^{-3} \frac{u_{*}^8}{u^8+u^5 u_{H}^3-u_{*}^8}, \qquad  u_{H}^3=R^3 {H}^2 .
\eea
For $u_{*}\neq 0$, the embeddings are U-shaped. Figure \ref{mag-owl-eight}(a) shows that, unlike for the case of a background electric field, there is just one U-shaped solution for a given value of the magnetic field.  As seen from Figure \ref{mag-owl-eight}(b), the minimum radius $u_t$ of a U-shaped brane increases with the magnetic field, which implies that the dynamical constituent mass of the fermions increases. Note that $u_t$ asymptotes to a fixed value at large $H$, which indicates that the fermion mass does the same.

To see how turning on a  background magnetic field affects  $\langle \OWL^j_i \rangle$ at zero temperature, we first consider the case where the (dimensionless) magnetic field $H$ is small. To leading order,  the flavor branes have a turning point at 
\bea
u_t \approx u_{*} \left[ 1-\frac{1}{8} \Big(\frac{u_{H}}{u_{*}}\Big)^{3} \right].
\eea
Also, the integral of (\ref{meomz}) can be expressed as 
\bea\label{ellmz}
\frac{\ell_0}{2}&=& R^{3/2} u_{*}^4 \int_{u_t}^{\infty} \frac{du}{ u^{3/2} \sqrt{u^8 +u^5 u_{H}^3-u_{*}^8}},\nonumber\\
&\approx& R^{3/2} u_{*}^4 \Big (I_3+ \frac{7}{16}u_{H}^3 I_9\Big).
\eea
Since $A_2$ does not couple to the worldsheet, the boundary conditions are not affected. Thus, the Euclidean worldsheet we considered in the previous sections remains a solution to the equations of motion. The area of the worldsheet in the small magnetic field approximation is given by
\bea\label{mactonzt}
S_{\rm F}&=&\frac{1}{2\pi \alpha^{'}} \int dx^4 \int_{u(x^4)}^{u_\Lambda} du,\nonumber\\
&\approx& \frac{\ell_0}{2\pi \alpha^{'}}u_\Lambda-\frac{1}{\pi \alpha^{'}} R^{3/2} u_{*}^4 \int_{u_t}^{\infty}  \frac{du}{u^{1/2}\sqrt{u^8-u_{cr}^3 u^5-u_{*}^8}},\nonumber\\  
&\approx& -\frac{1}{\pi \alpha^{'}} R^{3/2} u_{*}^4 (I_1+\frac{5}{16}\bar u_{H}^3 I_7),
\eea
where we have dropped the linear divergent term in the third line. Eliminating $u_{*}$ between (\ref{ellmz}) and (\ref{mactonzt}) results in 
\bea\label{acm}
S_{\rm F}=-c\lambda_{\rm eff}\left[1+a_H\Big(\frac{\ell_o}{R}\Big)^6 {H}^2\right],
\eea
where
\bea\label{defah}
a_H=2^8\Big[ B\Big( \frac {13}{16},\frac {1}{2} \Big)\Big]^{-6}
\left[5\frac{B\Big(\frac {13}{16},\frac {1}{2} \Big)}{B\Big( \frac {7}{16},\frac {1}{2} \Big)}+7
\frac{B\Big( \frac {15}{16},\frac {1}{2} \Big)}{B\Big( \frac {9}{16},\frac {1}{2} \Big)}\right]\approx 3.6,
\eea
and $c$ is defined in (\ref{defc}). Denoting the order parameter of the model when there is a background magnetic field by $\langle \OWL^j_i \rangle_{H}$, we have
\bea
\frac{\langle \OWL^j_i \rangle_{H}}{\langle \OWL^j_i \rangle}= 1+0.029 \lambda_{\rm eff}\Big(\frac{\ell_o}{R}\Big)^6 {H}^2.
\eea
Thus, for a small background magnetic field, the order parameter \emph{increases} quadratically with the magnetic field. Note that the above expression can also be obtained by taking $E^2\rightarrow -H^2$ in (\ref{owlE}).

\begin{figure}[h]
   \epsfxsize=2.5in \centerline{\epsffile{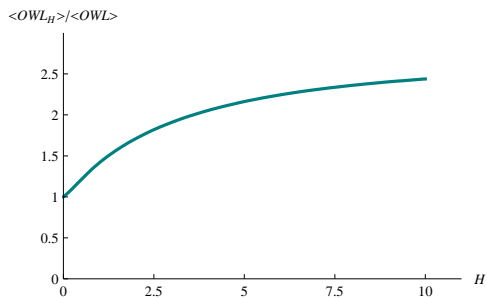}}
   \caption[FIG. \arabic{figure}.]{\footnotesize{At zero temperature, $\langle \OWL^j_i \rangle_{H}/\langle \OWL^j_i \rangle$ increases with the magnetic field, where for convenience we set $\ell_0=1$, $R=1$, $2\pi\alpha'=1$ and $u_\Lambda=10$.  }}
\label{owl-six}
\end{figure}

This result obtained for holographic NJL at strong coupling agrees with the magnetic catalysis of the standard NJL model in the presence of a background magnetic field \cite{gms9509}.  In fact, the authors of \cite{gms9509} have demonstrated that, not only is the critical coupling of the NJL model lowered by a background magnetic field, but the condensate increases quadratically in the limit of small magnetic field when the coupling is much larger than the critical coupling (see equation (58) in \cite{gms9509}).  In Figure \ref{owl-six}, we have set $R=1$, $\ell_0=1$, $2\pi\alpha^{\prime}=1$ and  $u_{\Lambda}=10$ and plotted $\langle \OWL^j_i \rangle_{H}/\langle \OWL^j_i \rangle$ versus magnetic field $H$ for generic values of the magnetic field. 

\subsubsection*{Finite temperature} 

We have already given a general argument for why, at high enough temperatures for which the theory is in the $\XSR$ phase, $\langle \OWL^j_i \rangle^\beta_{H}$ vanishes for any $H$. On the other hand, in the $\XSB$ phase with small temperature and small magnetic field, by taking $E^2\rightarrow -H^2$ in (\ref{owlET}) we find that
\bea
\frac{\langle \OWL^j_i \rangle_{H}^{\beta}}{\langle \OWL^j_i \rangle}= 1+1.09\lambda_{\rm eff} \Big(\frac{\ell_0}{\beta}\Big)^{6}+0.029 \lambda_{\rm eff}\Big(\frac{\ell_o}{R}\Big)^6 {{H}}^2.
\eea
Since the leading terms in $\beta^{-1}$ and $H$ do not mix, $\langle \OWL^j_i \rangle_{E}^{\beta} /\langle \OWL^j_i \rangle$ retains the same qualitative behavior that it had when either $\beta^{-1}$ or $E$ vanished.

\begin{figure}[h]
   \epsfxsize=6.0in \centerline{\epsffile{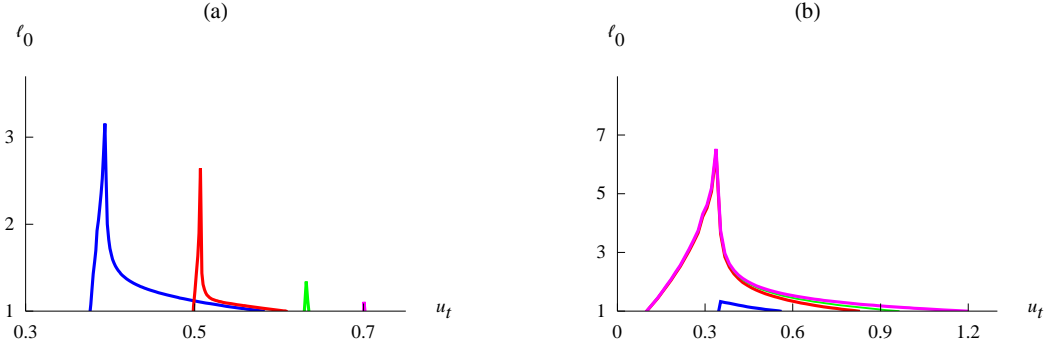}}
   \caption[FIG. \arabic{figure}.]{\footnotesize{ (a) $\ell_0$ versus $u_t$ for $H=0.1$ and various temperatures: $\beta^{-1}= 0.15$ (blue), $ 0.17$ (red), $ 0.19$ (green) and  $ 0.2$ (violet). We have set $R=1$. (b) $\ell_0$ versus $u_t$ for $\beta^{-1}=0.14$ and $H=0.05$ (blue), $1$ (red), $ 2$ (green) and $ 10$ (violet).}}
\label{mag-owl-nine}
\end{figure}

We will now use numerical techniques to consider how the vacuum of the model responds to simultaneously turning on arbitrary amounts of temperature and magnetic field. From the plots in Figure \ref{mag-owl-nine}, we see that turning on the temperature results in two U-shaped solutions, as in the case of vanishing magnetic field. In particular, Figure \ref{mag-owl-nine}(a) shows $\ell_0$ as a function of the turning point $u_t$ for $H=0.1$ and various temperatures. This demonstrates that there is a maximum temperature beyond which U-shaped solutions with $\ell_0=1$ cannot exist. A comparison of Figure \ref{mag-owl-nine} (a) to Figure \ref{owl-one} (a) shows that this maximum temperature is greater than the maximum temperature when there was no magnetic field.  Figure \ref{mag-owl-nine}(b) shows $\ell_0$ versus $u_t$ for a fixed temperature ($\beta^{-1}=0.14$) and different values of $H$. Note that the qualitative behavior of all of these plots remains the same for all allowed values of $\ell_0$. 
\begin{figure}[h]
   \epsfxsize=5.5in \centerline{\epsffile{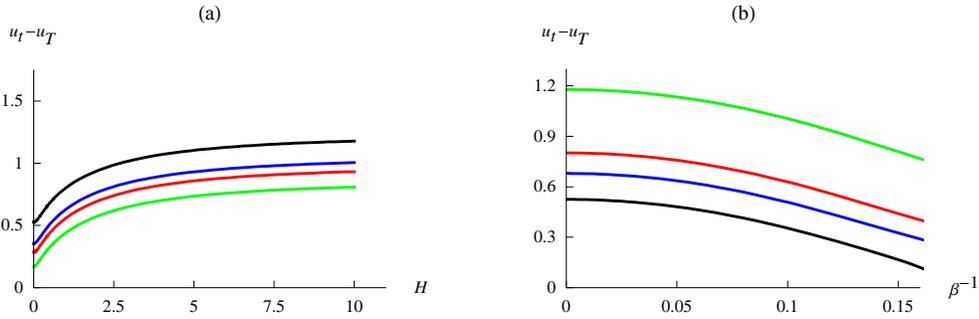}}
   \caption[FIG. \arabic{figure}.]{\footnotesize{ (a) $u_t-u_T$ versus $H$ for various temperatures: $\beta^{-1}= 0$ (black), $ 0.1$ (blue), $0.12$ (red) and $ 0.15$ (green). (b) $u_t-u_T$ versus $\beta^{-1}$ for $H=0$ (black), $ 0.5$ (blue), $1$ (red) and $ 0.10$ (green). The maximum temperature beyond which there is no U-shaped branes increases slightly with the magnetic field.}}
\label{mag-owl-ten}
\end{figure}
\begin{figure}[h]
   \epsfxsize=6in \centerline{\epsffile{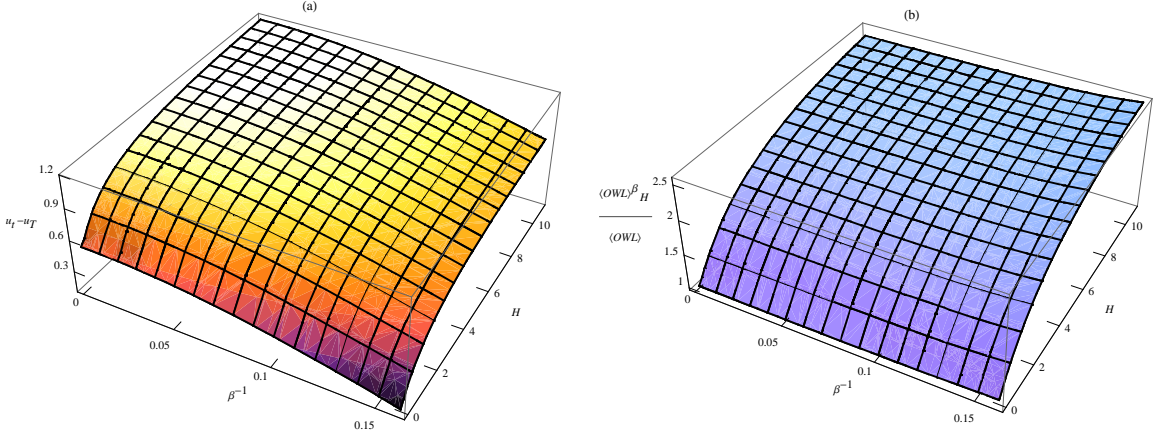}}
   \caption[FIG. \arabic{figure}.]{\footnotesize{ (a) $u_t-u_T$ as a function of $H$ and $\beta^{-1}$. (b) $\langle \OWL^j_i \rangle_{H}^{\beta} /\langle \OWL^j_i \rangle$ as a function of $H$ and $\beta^{-1}$.}}
\label{mag-owl-eleven}
\end{figure}

To explore how the turning point of the energetically-favored solution (with $\ell_0=1$ and $R=1$) depends on the temperature and magnetic field, we plotted $u_t-u_T$ versus $H$ for different temperatures in Figure \ref{mag-owl-ten} (a), and $u_t-u_T$ versus $\beta^{-1}$ for a few values of magnetic field in Figure \ref{mag-owl-ten} (b). Figure \ref{mag-owl-eleven} (a) shows a three-dimensional plot of $u_t-u_T$ as a function of $\beta^{-1}$ and $H$. This shows that $u_t-u_T$ decreases with $\beta^{-1}$, while it increases with $H$. 

The behavior of the $\langle \OWL^j_i \rangle_{H}^{\beta}$ as function of the temperature (of course, only up to the $\XSR$ temperature) and magnetic field is shown in Figure \ref{mag-owl-eleven} (b). $\langle \OWL^j_i \rangle_{H}^{\beta}$ monotonically increases with both $\beta^{-1}$  and $H$.

\section{Generalized OWLs and curved worldsheets}

According to Aharony and Kutasov's proposal \cite{ak0803}, in 
holographic NJL, $\langle \OWL^j_i \rangle\simeq \d^j_i e^{-S_{\rm F}}$ where $S_{\rm F}$ is the action of a Euclidean fundamental string whose worldsheet stretches between the regularized boundary $u=u_\L$ and the $\D8$-branes, whose ends are fixed to lie along the OWL contour. The string has Dirichlet boundary conditions along the contour at $u=u_\Lambda$ and Neumann boundary conditions on the D8-branes for $u<u_{\Lambda}$.  This worldsheet is, furthermore, assumed to lie at constant values of the other coordinate directions.  However, there exist other string worldsheets satisfying the OWL contour Dirichlet and D8-brane Neumann boundary conditions which do {\it not} lie at constant transverse coordinate values. The OWL and D8-brane boundaries force the string to be extended in the $u$- and $x^4$-directions, but we can also allow it to bend in, say, the $t$- or $x^{1,2,3}$-directions as well.  

In this section we will show the existence of such curved worldsheets and then discuss what is the corresponding holographic $\OWL$ operator.  We will concentrate on the $\XSR$ phase with parallel $\D8$- and $\overline{\D8}$-branes, since it is relatively easy to solve for curved worldsheets in this configuration.  However, in principle, one should be able to do a similar analysis for the worldsheets bounded by U-shaped branes in the $\XSB$ phase.  In what follows, we perform the computations both in Minkowski and Euclidean signatures of the background geometry using Schwarzchild or Kruskal coordinates, as appropriate. 

\subsection{Lorentzian signature---Schwarzchild coordinates} 

We start with Schwarzchild coordinates and work with a form of the background metric in which the coordinates are dimensionless, as we did in (\ref{nodimbhmetric2}). 

\subsubsection*{$x^0$-bending solutions}

We look for possible solutions which bend in the $x^0$- (or $t$-) direction.  We choose the gauge and embedding
\be
x^4=\s^1, \qquad u=\s^2,\qquad x^0 = t(u).
\ee
Then
the Nambu-Goto action is
\be
{S_{\rm F}}= \frac{\gamma^2\ell_0}{2\pi\alpha'} \int^{u_\L} du \sqrt{u^3-(u^3-1)^2(t')^2 \over u^3-1}.
\ee
The lagrangian is invariant under constant shifts in $t$, giving rise to a conserved charge density (energy) carried by the string.  So the first integral of the equation of motion is
\be\label{x01m}
{dt\over du} = { c_0\, u^{3/2} \over (u^3-1) \sqrt{u^3 - 1 + c_0^2} },
\ee
where $c_0$ is a constant.

The right side of (\ref{x01m}) has an integrable singularity at $u^3=1-c_0^2$ (corresponding to a turning point) and a non-integrable one at the horizon, $u=1$ (corresponding to an asymptote).  Thus, since solutions must start at $u=u_\Lambda >1$, they all asymptote to the horizon, since the turning point is hidden behind it. The only exception is $c_0=0$, for which constant $t$ corresponds to the straight string solution of Aharony and Kutasov.

For these solutions, the action becomes
\be
{S_{\rm F}}
\propto \int^{u_\L} {u^{3/2} du \over\sqrt{u^3 - 1 + c_0^2}}.
\ee
This action is integrable at the horizon, turning point and singularity, and is therefore finite unless the string goes off to $u=\infty$.  In particular, the solutions which asymptote to the horizon have finite area, and thus must continue through the horizon in some appropriate in-falling coordinate system, like Eddington-Finkelstein coordinates, which are regular near the horizon.  As this is just a change of variables from Schwarschild coordinates, the equation (\ref{x01m}) stays the same;  the only difference is that the coordinate change allows us to connect it smoothly to a solution with $u<1$ with the same value of $c_0$.  Such $u<1$ solutions asymptote to the horizon and reach the  integrable turning point ($u^3=1-c_0^2$).  What is less clear in these coordinates is what happens to the worldsheet after reaching this point.  To understand this, it is best to go to the maximally extended Kruskal coordinates, which we will do in the next subsection.

\subsubsection*{$\vec x$-bending solutions}

Now let's look for possible solutions which also bend in, say, the $x^2$- (or $y$-) direction.  Choose the gauge and embedding
\be
x^4=\s^1, \qquad u=\s^2,\qquad x^2 = y(u),
\ee
so that the action becomes
\be
{S_{\rm F}}= \frac{\gamma^2\ell_0}{2\pi\alpha'}\int^{u_\L} du\,u^{3/2}
\sqrt{{1\over u^3-1} + (y')^2 }.
\ee
The Lagrangian is invariant under constant shifts in $y$, giving rise to a conserved charge density (momentum in the $x^2$-direction) carried by the string.  So the first integral of the equation of motion gives
\be\label{yeq}
{dy\over du} =
{c_2 \over \sqrt{(u^3 - c_2^2 ) (u^3-1)} },
\ee
where $c_2$ is a constant.  When $c_2=0$, this is again the straight string solution of Aharony and Kutasov.  

For these solutions, the action is
\be
S_F\propto \int^{u_\L} 
{ u^3 du \over \sqrt{(u^3-1)(u^3 - c_2^2)} },
\ee
which is integrable both at the horizon ($u=1$) and the turning point
($u^3=c^2_2$), except when they coincide.
For $0<|c_2|<1$, the string reaches the horizon at finite $y$ and then turns back up.  Again, Kruskal coordinates are needed to understand the behavior of the string after this turning point, and will be discussed below.  For $|c_2|=1$, the string asymptotes to the horizon. Since it has infinite area, the string does not cross the horizon, even in in-falling coordinates.  For $1<|c_2|$, the string reaches a minimum radius of $u = |c_2|^{2/3}$ at finite $y$ and then turns back up.  These turning solutions are non-physical (or, infinite-area) since they have no $\OWL$ insertion to end on.  

\subsection{Lorentzian signature---Kruskal coordinates} 

The Kruskal coordinates have been defined in (\ref{kruscoor}) in terms of which the background metric was given in (\ref{actionkrus}).  Here we use them to analyze the $x^0$- and $\vec x$-bending worldsheets.  However, first recall that the straight (Aharony-Kutasov) worldsheet which was the main focus of the previous sections corresponds to the line $w = -e^{-3t_0} v$.  This worldsheet, depicted by the green line in Figure \ref{owl-krusk2}, passes into the second asymptotic region, where it has nowhere to end since there is no $\OWL$ insertion at $u=u_\L$ there for it to end on. Such a worldsheet presumably would contribute to a finite temperature two-point function of OWLs.

\subsubsection*{$x^0$-bending solutions}

\begin{figure}[h]
   \epsfxsize=4in \centerline{\epsffile{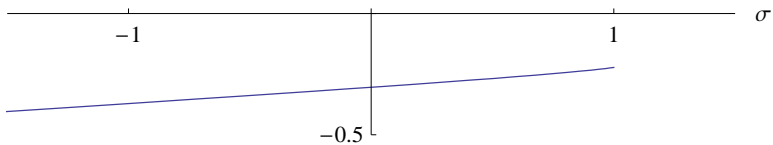}}
   \caption[FIG. \arabic{figure}.]{\footnotesize{The square bracket in the fundamental string Lagrangian as a function of $\s$.}}
\label{owl-krusk3}
\end{figure}

Choose the gauge and embedding
\bea
x^4=\s^1, \qquad vw=\s^2:=\s,\qquad v/w = \t(\s),
\eea
so that the action is given by
\be
S_{\rm F}= \frac{\gamma^2\ell_0}{2\pi\alpha'}\int^1_{\s_\L} d\s \sqrt{\left[{1\over9}\cdot{u^3-1\over - e^{3r}}\right] {1-\s^2 (\t'/\t)^2\over \s}}.
\ee
Here, $\s_\L$ is related to $u_\L$ by $-e^{3r(u_\L)} := \s_\L$, and is negative.  Since $u$ and $r$ are definite functions of $vw$, the quantity in square brackets is a definite (known) function of $\s$.  In fact, it depends only mildly on $\s$, as shown in Figure \ref{owl-krusk3}.

\begin{figure}[h]
   \epsfxsize=3in \centerline{\epsffile{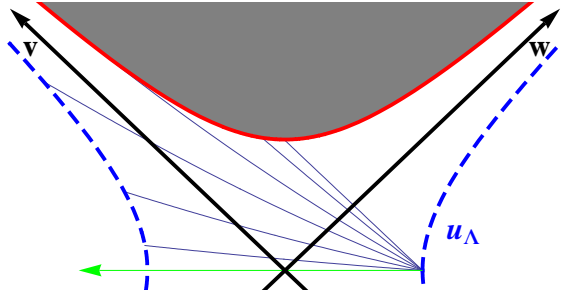}}
   \caption[FIG. \arabic{figure}.]{\footnotesize{The gray lines are string worldsheets for $c_0=0.1$, $0.3$, $0.6$, $1$, $2$, and $10$ in ascending order from the green $c_0=0$ line (green).}}
\label{owl-krusk4}
\end{figure}

In Schwarschild coordinates, the Lagrangian is invariant under time translations: $t\to t+\mbox{constant}$.  This translates to a scaling invariance under $\t \to \t \cdot e^{-3\cdot\mbox{constant}}$ (since $\t=-e^{-3t}$).  The first integral of the corresponding equation of motion gives
\bea\label{x02}
{1\over\t}{d\t\over d\s} =
{-c_0 \over \s \sqrt{(u^3 - 1+c_0^2 )} }.
\eea
As a check, upon changing variables from $\s$ to $u$, the above expression becomes identical to (\ref{x01m}).  Thus, we can simply integrate (\ref{x01m}) (on both sides of the horizon and of the turning point, if the string gets there) and join the solutions by changing to Kruskal coordinates using the above formulae.  This needs to be done numerically, giving the result shown in Figure \ref{owl-krusk4}.  The fact that the worldsheets look like straight lines is misleading. In actuallity, they deviate very slightly from straight lines, essentially by the logarithm of the function shown in Figure \ref{owl-krusk3}. If one had chosen a larger value of $u_\L$ (that is, larger than the choice of $u_\L=1$ which is shown), then the curvature would be apparent.  There is no good reason for ending the $|c_0|<1$ lines at $u=u_\L$ in the second asymptotic region.  The $c_0=1$ line is just tangent to the singularity.  The $c_0\to\infty$ line becomes lightlike {\footnote{It is not clear that worldsheets which hit the singularity (region of high curvature) can be trusted. It may be possible for such worldsheets to avoid the singularity by following complexified geodesics to the second asymptotic region \cite{fhks0306}.}}.  

\subsubsection*{$\vec x$-bending solutions}

A similar analysis can be done for the $y$-bending solutions.  The behavior of these solutions in the global Kruskal coordinates is easy to deduce from what we already know.  The $c_2=0$ solution is the straight string which extends into the second asymptotic region.  Likewise, all the $0<|c_2|<1$ solutions extend to infinity in the second asymptotic region.  The $|c_2|=1$ solutions asymptote to the horizon and extend to $|y|=\infty$.  Finally, the $|c_2|>1$ solutions turn at a finite altitude above the horizon and extend back to infinity in the first asymptotic region.  Since none of these solutions bend in the $x^0$-direction, none of them hit the singularity.

\subsection{Euclidean signature} 

We now do the same analysis in Euclidean signature.  Since the $\OWL$  is at a fixed time, the Euclidean continuation of its one-point function is an insertion at fixed Euclidean time $t_E$ and should be interpreted as a finite temperature expectation value of the OWL operator.

The Euclideanized metric is just (\ref{nodimbhmetric2}) with $dt^2\to-dt_E^2$
\be\label{Ebhmetric}
\frac{ds_E^2}{\gamma^2} = {u^3-1 \over u^{3/2}} dt_E^2 
+ u^{3/2} d{\vec x}^2
+ {u^{3/2}\over u^3-1} du^2 + u^{1/2} d\Omega_4^2 .
\ee
Near $u=1$, this looks approximately like
\be
\frac{ds^2} {\gamma^2}\approx 
3(u-1) dt_E^2 + d{\vec x}^2
+ {1\over 3(u-1)} du^2 + d\Omega_4^2 
= d\r^2 + \r^2 d\theta^2 + d{\vec x}^2
+ d\Omega_4^2, 
\ee
where in the second equality we changed coordinates $u\to\r:=2\sqrt{u-1}/\sqrt3$ and $t_E\to\theta:=3t_E/2$, since $t_E$ has period $4 \pi/3$ for the geometry to be smooth at $u=1$.

\subsubsection*{$x^0$-bending solutions}

\begin{figure}[h]
   \epsfxsize=3in \centerline{\epsffile{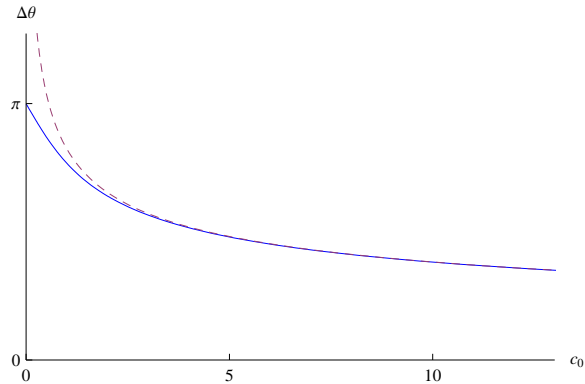}}
   \caption[FIG. \arabic{figure}.]{\footnotesize{$\Delta\theta$ as a function of $c_0$. The dashed line is $(6 \sqrt\pi \Gamma[{2\over3}])/
(c_0^{1/3} \Gamma[{1\over6}])$, which is the asymptotic form of $\Delta\theta(c_0)$.}}
\label{owl-krusk5}
\end{figure}

With the gauge and embedding
\be
x^4=\s^1, \qquad u=\s^2,\qquad x^0 = t_E(u),
\ee
one finds that the action is given by
\be
{S_{\rm F}}
= \frac{\gamma^2\ell_0}{2\pi\alpha'} \int^{u_\L} du \sqrt{u^3+(u^3-1)^2(t_E')^2 \over u^3-1}.
\ee
From this, we find
\be\label{x01}
{dt_E\over du} = 
{ c_0\, u^{3/2} \over (u^3-1) \sqrt{u^3 - 1 - c_0^2} }.
\ee
For $c_0\neq0$, solutions all reach the turning point at $u^3=1+c_0^2 >1$ at finite (angle) $t_E$, at which point the solution then goes back to large $u$ at some new asymptotic value of $t_E$.  There is no value of $c_0$ for which the behavior changes qualitatively.  At $c_0=0$, $t_E=\mbox{constant}$ down to $u=1$.   Near $u=1$, this simply means that it is a straight line in the $\{\r,\theta\}$ polar coordinates approaching the origin at a constant value of $\theta$, say $\theta=0$.  Thus, it continues through the origin coming out at $\theta=\pi$. Figure \ref{owl-krusk5} shows the value of $\Delta\theta$, the difference in the asymptotic values of $\theta = 3t_E/2$ for large $u$, as a function of $c_0$.  Since all of these solutions turn and head back out to large $u$ where there is no $\OWL$ operator insertion for them to end on, they should be discarded as solutions (or considered to have infinite area).

\subsubsection*{$\vec x$-bending solutions}

As before, we choose the gauge and embedding $x^4=\s^1$, $u=\s^2$ and $x^2 = y(u)$. Since the time coordinate does not come into play anywhere, the result is identical to the previously-discussed Lorentzian case, with the same qualitative behavior of the string worldsheets. Namely, for $|\vec c|<1$ the string goes through the $u=1$ origin and comes out at $\Delta\theta=\pi$ (and a different value of $\vec x$ as well). For $|\vec c|=1$ the string never makes it to $u=1$ but instead asymptotes to $u=1$ while extending infinitely along a $\vec x\propto\vec c$ direction. For $|\vec c|>1$ the string stays at $\theta=0$ and turns back up in $u$ before reaching $u=1$.

\subsection{Interpretation of the curved worldsheet solutions} 

The integration constants $c_\mu$ that appeared above are adjustable properties of the solutions, and should therefore reflect a change in the operator whose one-point function is being evaluated.  Thus, instead of a unique operator $\OWL^i_j(x^\mu)$, as defined by Aharony and Kutasov, there must be an entire family of similar operators, $\OWL^i_j(x^\mu;c_\mu)$.  What are these generalized OWL operators in holographic NJL?

From the gravity side, it is not too hard to see what they are.  Non-zero $c_\mu$ imply that the string worldsheet no longer ends perpendicularly on the Wilson line at $u=u_\Lambda$.  Equivalently, the $c_\mu$ are proportional to non-zero fluxes of spacetime energy-momentum flowing into the string.  This means that there are forces on the ends of the string.

Since strings end on D-branes, in order for there to be a precise string dual of the Aharony-Kutasov OWL operator, it should be thought of as a contour on an appropriate probe D4- or D6-brane (at $u=u_\Lambda$ and parallel to the $N_c$ color D4-branes) on which the string ends.  In this case, the source of the forces on the string endpoints when $c_\mu\neq0$ is apparent;  they are constant electromagnetic fields $F_{\mu\nu} \propto \d_{4[\nu}c_{\mu]}$ on the probe D4- or D6-brane\footnote{This situation is similar to that of time-like strings in the D3-D7 system, for which the Wilson line is the path of the string endpoint on the D7 brane, while the integration constants are interpreted as forces applied to the ends of the string by constant electromagnetic fields on the D7 brane.  In the present case, we are essentially describing the same thing except with spacelike, instead of timelike, Wilson lines.}.

Some interesting things emerge from this picture.  In the Lorentzian case, turning on $c_0$ corresponds to an electric field on the probe D6-brane.  $|c_0|=1$ corresponds to the critical electric field beyond which the D6-brane no longer exists, and such an operator cannot be defined.  We have seen that the corresponding $|c_0|\ge 1$ string worldsheets are those that hit the black hole curvature singularity.  Likewise, turning on $\vec c$ corresponds to a magnetic field on the D6-brane.  Here, there is no critical value but one expects that, for $|\vec c|>1$, there is a qualitative change in behavior as the strong magnetic field acts to ``confine" the D6-brane to a smeared D4-brane; indeed, a qualitative change happens here---the strings go through to the second asymptotic region for $|\vec c|<1$ but turn and go back out the first asymptotic region when $|\vec c|>1$.

However, this pleasing picture on the string side does not shed much light on the description of the $c_\mu\neq0$ OWL operators in the effective five-dimensional Yang-Mills plus four-dimensional fermion gauge theory.  Since the D6-brane electromagnetic fields are constant, the modification to the Aharony-Kutasov OWL operator should be distributed uniformly along the length of the five-dimensional Wilson line.  It presumably corresponds to ``dressing" this Wilson line with a density of gauge covariant insertions.  One possibility, which breaks the four-dimensional Lorentz invariance in the correct way is
\be
\OWL_i^j(x^\mu,c^\nu)  \,\stackrel{?}{=} \,
\psi^{\dagger j}_L(x^\mu,-\ell_0/2)\,\,
{\cal P}\!\ \exp\left[\int_{-\ell_0/2}^{\ell_0/2} (iA_4+c^\nu F_{\nu 4}+\Phi) dx^4\right] 
\psi_{iR}(x^\mu,\ell_0/2) .
\ee

In any case, whatever their precise form, their expectation value is a $\XSB$ order parameter.  We have found that, in the parallel D8- and $\overline{\D8}$-brane geometry, all the corresponding string worldsheets have infinite area and thus lead to a vanishing order parameter, consistent with being in the $\XSR$ phase.

\section{More on generalized OWLs and curved worldsheets}

In this section, we consider more general worldsheets which are dual to the $\OWL$ operators in 
which the fermions are located at two different points in the $x^{\mu}$-directions. We first consider the case 
in which the left and the right-handed fermions are separated only in the $x^0$-direction, and then analyze the case in which they are separated in the $\vec{x}$-directions.

\subsection*{$x^0$-bending solutions}

Consider the following gauge and embedding 
\be
x^4=\s^1, \qquad u=\s^2, \qquad x^0=t(x^4, u).
\ee
Following \cite{ak0803}, this embedding should describe  a worldsheet whose holographic dual is  an $\OWL$ operator with chiral fermions separated in the $x^0$-direction. The worldsheet has Neumann boundary conditions along the directions of $\D8$ and $\overline{\D8}$-branes and Dirichlet boundary conditions for directions normal to them. With the above choice of gauge and embedding, the worldsheet action reads
\bea
S_{\rm F}=\frac{\gamma^2}{2\pi\alpha^{'}}\int dudx^4 \sqrt{G_{uu}G_{44}+  G_{tt}G_{44} (t')^2 + G_{tt}G_{uu} (\dot t)^2},
\eea
where in this section we use $^.$ and $'$ to denote derivatives with respect to $x^4$ and $u$, respectively. For the worldsheet to be well-defined everywhere, one has to impose the condition
\bea\label{cond}
\frac{1}{\gamma^2}\hbox{det} (h_{\alpha\beta})=G_{uu}G_{44}+  G_{tt}G_{44} (t')^2 + G_{tt}G_{uu} (\dot t)^2>0.
\eea

The Lagrangian is invariant under a shift in $x^0$, giving rise to the equation of motion
\be\label{par}
\left(\frac{G_{tt} G_{uu} {\dot t}}{\sqrt{G_{uu}G_{44}+  G_{tt}G_{44} (t')^2 + G_{tt}G_{uu} (\dot t)^2}}\right)^{.} +
\left(\frac{G_{tt} G_{44} {t'}}{\sqrt{G_{uu}G_{44}+  G_{tt}G_{44} (t')^2 + G_{tt}G_{uu} (\dot t)^2}}\right)' =0.
\ee
Although we do not know a closed form solution to (\ref{par}), its solution can be obtained numerically.  

However, instead of following this route, we will restrict our ansatz further in the hope of finding an analytic solution to the equation of motion.  Choosing
\bea\label{resansatz}
x^0= k x^4 + t(u),
\eea
with $k$ being a constant, drastically simplifies the equation of motion.  Indeed, substituting (\ref{resansatz}) into (\ref{par}) results in an the following first integral of the equation of motion
\bea\label{simeom}
\frac{G_{tt} G_{44} {t'}}{\sqrt{G_{uu}G_{44}+G_{tt}G_{uu} k^2+  G_{tt}G_{44} (t')^2} }=c,
\eea
where $c$ is a constant of integration (not to be confused with $c$ defined in (\ref{defc})). 

Note, however, that the simple ansatz of (\ref{resansatz}) does not  solve the Neumann boundary conditions for the worldsheet. One can consider the contour of (\ref{resansatz}) as a good approximation to the contours where the complications due to the edges can be ignored. In that sense, the ansatz (\ref{resansatz}) merits further study. 
Solving (\ref{simeom}) for $t'$ results in
\bea\label{tpri}
t'^2 =c^2\frac{u^3-k^2(u^3-1)}{(u^3-1)^2(c^2+u^3-1)},
\eea
and substituting (\ref{tpri}) into  (\ref{cond}) yields
\bea\label{co}
\frac{1}{\gamma^2}\hbox{det} h_{\alpha\beta}
=\frac{u^3-k^2(u^3-1)}{c^2+u^3-1}>0.
\eea

The positivity of the determinant of the induced metric (\ref{co}), together with the reality of the solution (\ref{tpri}), implies that $c^2\geq 0$. Now, $c=0$ does not represent a physical solution for the following reason.   At $u_{*}^3=k^2/(k^2-1)$ the determinant changes sign, implying that the allowed range of $u$ is $[u_{\Lambda}, u_{*})$. Therefore, for $c=0$ to be a solution, the worldsheet must either have a turning point at some $u_{t} \geq u_{*} $ so that it never reaches $u_{*}$ or, if it does not have turning points, then it must end at $u=u_{*}$.  Equation (\ref{tpri}) implies that the only turning point is at the horizon $u=1$, which is actually below $u_{*}$. Thus, the solution must end at $u=u_{*}$ before it hits the would-be turning point. However, there is nothing for the worldsheet to end on as there are no D-branes there.  Therefore, we discard the $c=0$ branch as a physical solution.  
 
For $c^2>0$, the story is the same as the $c=0$ case. In order to obtain a valid worldsheet,  (\ref{co}) indicates that again the worldsheet must end at $u=(k^2/(k^2-1))^{1/3}$.  Since there is no D-brane there for the worldsheet to end on, we discard these solutions as well.  So we conclude that, modulo a subtlety having to do with the boundary conditions, since there is no classical worldsheet of the type (\ref{resansatz}) the vev of the dual holographic operator must vanish in the $\XSR$ phase of the model.  It is reassuring  that the vev of this operator vanishes in the chirally-symmetric phase because we argued above that the dual operator is a chirally-charged OWL where the left-handed and the right-handed fermions are separated in the $x^0$-direction. 

\subsection*{$\vec x$-bending solutions}

For an OWL with the left and the right-handed fermions separated in, for example, the $y$-direction, we choose the following gauge and embedding for the holographic worldsheet
\bea\label{xan}
x^4=\s^1, \qquad u=\s^2, \qquad y=y(x^4, u). 
\eea
The worldsheet action takes the form
\bea\label{xact}
S_{\rm F}=\frac{\gamma^2}{2\pi\alpha^{'}}\int dudx^4 \sqrt{G_{22} \Big[G_{uu}\Big(1+(\dot y)^2\Big) + G_{22}(y')^2\Big]}.
\eea
The Euclidean worldsheet must satisfy the condition
\bea
G_{22} \Big[G_{uu}\Big(1+(\dot y)^2\Big) + G_{22}(y')^2\Big]>0.
\eea
Since the integrand in (\ref{xact}) is independent of $y$, the equation of motion is
\bea\label{pa}
\left(\frac{ G_{22} G_{uu} {\dot y}}{\sqrt{G_{22} \Big[G_{uu}\Big(1+(\dot y)^2\Big) + G_{22}(y')^2\Big]}}\right)^{.} +\left(\frac{ (G_{22})^2 {y'}}{\sqrt{G_{22} \Big[G_{uu}\Big(1+(\dot y)^2\Big) + G_{22}(y')^2\Big]}}
\right)' =0.
\eea

In order to find a closed-form solution, we relax (\ref{xan}) and choose a simpler ansatz
\bea\label{yan}
x^4=\s^1, \qquad u=\s^2, \qquad y=k x^4+ y(u). 
\eea
Again, this simple ansatz does not satisfy the Neumann boundary conditions but is still worth studying because many contours can be approximated by such an ansatz as long as the edges of the worldsheet do not drastically change the shape of the worldsheet. 

Substituting (\ref{yan}) into (\ref{pa}) yields the first integral of motion
\bea\label{yp}
\frac{ (G_{22})^2 {y'}}{\sqrt{G_{22} \Big[G_{uu}(1+k^2) + G_{22}(y')^2\Big]}} =c,
\eea
where $c$ is a constant of integration.  Solving (\ref{yp}) for $y'$ and substituting the values of $G_{22}$ and $G_{uu}$ gives
\bea\label{ypf}
y'^2= (1+k^2)c^2 \frac{1}{(u^3-1)(u^3-c^2)}.
\eea
Except for a factor of $1+k^2$, (\ref{ypf}) is identical to (the square of) (\ref{yeq}). Thus, our analysis in the previous section goes through here as well. In particular, when $c=0$, we have a straight worldsheet which passes through the horizon and hits the second asymptotic boundary. Since there is no OWL insertion at the second asymptotic boundary, this worldsheet solution has nowhere to end and so one should discard this solution. When $c^2=1$, the worldsheet asymptotes to the horizon. The area of the worldsheet is 
\bea
S_{\rm F}= \frac{\ell \gamma^2}{2\pi\alpha^{'}} \sqrt{1+k^2}\int_1^{u_\Lambda} du \frac{u^3}{u^3-1},
\eea
which is infinite (besides the usual UV-divergent piece) giving zero contribution to the one-point function of the dual $\OWL$ operator. When $0<c^2<1$, the worldsheet dips into the horizon and has a turning point at some radial position below the horizon from which it turns back up hitting the second asymptotic boundary of the spacetime.  Since there is no OWL insertion on the second asymptotic boundary for the worldsheet to end one, one should discard this solution as well.  Finally, when $c^2>1$, the worldsheet has a turning point at  $u_t = |c_2|^{2/3}$ (above the horizon) and turns back up to the boundary of the spacetime for which there is no $\OWL$ insertion to end on. This solution is non-physical too.  Therefore, one reaches the conclusion that, in the $\XSR$ phase of holographic NJL, the one-point functions of the $\OWL$s described above (those in which the fermions are separated in the $y$-direction) all correctly vanish. 

\section*{Acknowledgments}

We would like to thank O. Aharony, S. Baharian, O. Bergman, M. Kruczenski, S. Mathur, V. Miransky, P. Ouyang and R. Wijewardhana for helpful discussions.  We would also like to thank the Aspen Center for Physics for hospitality where this work was initiated.  R.G.L.\ and M.E.\ are supported by DOE grant FG02-91-ER40709 and P.C.A.\ is supported by DOE grant FG02-84-ER40153.

\end{document}